\documentclass{aastex}
\usepackage{emulateapj5}
\usepackage{apjfonts}
\usepackage{epsf}

\usepackage{color}

\shorttitle{Abundance patterns in the ISM of early-type galaxies}
\shortauthors{Konami et al.}

\begin{document}

\title{Abundance Patterns in the Interstellar Medium of early-type
galaxies observed with {\it Suzaku}}

\author{SAORI KONAMI\altaffilmark{1,2} , KYOKO MATSUSHITA\altaffilmark{2}, RYO NAGINO\altaffilmark{3}, and TORU TAMAGAWA\altaffilmark{4,2}}

\altaffiltext{1}{Department of Physics, Tokyo Metropolitan University,
1-1 Minami-Osawa, Hachioji, Tokyo 192-0397}
\email{konami@tmu.ac.jp}
\altaffiltext{2}{Department of Physics, Tokyo University of Science,
1-3 Kagurazaka, Shinjuku-ku, Tokyo 162-8601}
\altaffiltext{3}{Department of Earth and Space Science, Graduate School of Science, 
Osaka University, Toyonaka, Osaka 560-0043}
\altaffiltext{4}{High Energy Astrophysics Laboratory, 
RIKEN (The Institute of Physical and Chemical Research),
2-1 Hirosawa, Wako, Saitama 351-0198}

\begin{abstract}
We have analyzed 17 early-type galaxies, 13 ellipticals and 4 S0's,
 observed with {\it Suzaku}, and investigated 
metal abundances (O,  Mg, Si, and Fe) and abundance ratios (O/Fe,  Mg/Fe, and Si/Fe) 
in the interstellar medium (ISM). 
The emission from each on-source region, which is 4 times 
 effective radius, $r_{\rm e}$,
  is reproduced with  one- or two- temperature 
thermal plasma models as well as a multi-temperature model, using  APEC plasma code v2.0.1. 
The multi-temperature model gave almost the same abundances and abundance ratios 
with the 1T or 2T models.
The  weighted  averages of the O,  Mg, Si, and Fe abundances of all the
 sample galaxies derived from the multi-temperature model fits are 0.83 $\pm$ 0.04, 
0.93$\pm$0.03, 0.80 $\pm$ 0.02, and 0.80$\pm$0.02 solar, respectively, 
in solar units according to the solar abundance 
table by \citet{lodders_03}. 
These abundances
show  no significant dependence on the morphology and environment.
The systematic differences in the derived  metal abundances between the
 version 2.0.1 and 1.3.1 of APEC plasma codes were investigated.
The derived O and Mg abundances in the ISM agree with the stellar metallicity within
 a aperture with a radius of  one $r_{\rm e}$ derived from optical spectroscopy.
From these results, 
we discuss the past and present SN Ia rates and star formation histories
in early-type galaxies.

\end{abstract}


\section{INTRODUCTION}
\label{sec:intro}

The study on 
formation and evolution of early-type galaxies is one  of the most
important topics  in  modern astrophysics. 
The metal abundances of stars in these galaxies give us important constraints
on theoretical models of their formation.
In early-type galaxies, stellar mass loss and present type Ia supernovae (SNe Ia) 
have been providing metals in hot X-ray emitting interstellar medium (ISM).
Thus, the metal abundance and their ratios in the ISM
can provide a cumulative fossil record on the history of star formation.

\citet{Matsushita_01} found that
the X-ray luminosities in normal early-type galaxies which are not
located at the center of group or cluster scale potential are consistent
with the energy input from stellar mass loss.
As a result,  the hot ISM in these galaxies reflects almost instantaneous
balance between  heating and cooling.
The mass of  X-ray emitting ISM in early-type galaxies within 4 times 
the effective radius, $r_{\rm e}$,
 are about 0.1--1\% of the stellar mass \citep{Matsushita_01}.
The  timescale for accumulation of this amount of hot ISM 
is smaller than $\sim$ 1 Gyr.
Therefore, metal abundances in the ISM can constrain the present metal
supply by SN Ia and stellar mass loss.

The {\it ASCA} satellite  enabled us to measure the metal abundances 
in the ISM of  early-type galaxies through spectral fitting
of Fe-L lines.
Since  Fe in the ISM comes from stellar mass loss and SNe Ia,
the Fe abundance in the ISM is expected to  be a  sum of the stellar metallicity and
the contribution from SNe Ia, which is proportional to the ratio of 
SN Ia rate to stellar mass loss rate (see \cite{matsushita_03} for
details).
Adopting the SN Ia rate recently estimated with optical observations,
the resultant Fe abundance from SN Ia is at least 2 solar
\citep[e.g.,][]{arimoto_97, matsushita_03, nagino_10, konami_10,  loewenstein_10}.
However, the early measurements of ISM with ASCA showed that the 
metallicity is less than half a solar abundance
 \citep[e.g.,][] {awaki_94,loewenstein_94, mushotzky_94, matsushita_94}. 
Later, using the same ASCA data, 
the derived metal abundances in the ISM of these early-type galaxies have
 become $\sim$ 1 solar, 
considering uncertainties in the Fe-L atomic data \citep[]{arimoto_97, matsushita_97,  matsushita_00}, 
or employing a multi-temperature plasma model \citep{buote_98}. 
Using plasma codes with revised atomic data for the Fe-L lines, 
 the  Reflection Grating Spectrometer (RGS) onboard {\it XMM-Newton}, 
and the CCD detectors onboard {\it Chandra} and {\it XMM-Newton}
yielded  the Fe abundances in the ISM of  $\sim$ 1 solar with a significant scatter.
\citep[e.g.,][]{xu_02, werner_09, humphrey_06, werner_06, tozuka_08, ji_09}. 
These Fe abundances are comparable to the stellar metallicity 
of these galaxies \citep{arimoto_97, kobayashi_99}, but 
still smaller than the expected values from  SN Ia.
\citet{arimoto_97} have discussed various astrophysical aspects to
the low Fe abundances in the ISM.
\citet{matsushita_00} suggested  that SN Ia products in the ISM in 
early-type galaxies are lost to 
intergalactic space by their buoyancy. 
\citet{tang_10} also simulated the evolution of hot SN Ia ejecta, 
and found that they quickly reach 
a substantially higher outward velocity than the ambient medium. 
\citet{loewenstein_10} also suggested the "effective" rate of SN Ia enrichment is less than 
the actual rate because SN Ia is not efficiently well-mixed into the ISM. 
There are, however, no clear theoretical or observational evidence which can resolve this discrepancy.

Since present star formation activity
in early-type galaxies are generally low  and SNe Ia rarely produce O or Mg, 
abundances of theses elements in the ISM 
should reflect stellar abundances. 
Using RGS detectors,
the K$\alpha$ lines of O, Ne, and Mg from central regions of 
 X-ray luminous early-type
 galaxies were clearly detected \citep{xu_02, ji_09, werner_09}.
The derived metal abundances of O, Ne, and  Mg  were about 0.5--1  solar.
Using CCD detectors onboard {\it Chandra} and {\it XMM}, the abundance ratios such as
O/Fe, Mg/Fe and Si/Fe were also derived \citep{humphrey_06, tozuka_08}.
The derived Mg/Fe and Si/Fe  ratios are mostly consistent with the solar ratio,
while the O/Fe ratios are often significantly smaller than the solar ratio.
The {\it Suzaku} XIS \citep{koyama_07} has a good energy resolution at the O line 
energy, with a lower and more stable background level compared to the 
CCD detectors onboard {\it Chandra}  and  {\it XMM-Newton} and 
 a larger effective area compared to the RGS detector.
These advantage of {\it Suzaku} XIS can
reduce the statistical and systematic errors in O and Mg abundance measurements.
With {\it Suzaku},   the O, Mg, Si, and Fe 
abundances in the ISM in the entire region of  several elliptical galaxies
have been measured
 \citep[e.g.,][]{matsushita_07, tawara_08, hayashi_09,
loewenstein_10, loewenstein_12}. 
The derived  O/Fe,  Mg/Fe, and Si/Fe ratios 
are close to that of the new solar abundance ratios determined by
\citet{lodders_03}.
{\it Suzaku} also measured the metal abundances in the ISM  of 
two S0 galaxies, NGC~4382 and NGC~1316 \citep[]{nagino_10, konami_10}. 
The O/Fe ratio in the ISM in NGC~4382  is smaller by a factor of two than 
those of the other early-type galaxies.

The stellar metallicity reflects the past activity of star formation.
In addition,  a  longer formation time provides a higher concentration
of trapped SNe Ia products to the stars.
The stellar metallicity of the early-type galaxies has often been 
investigated in optical observations
of their central regions using Mg and Fe absorption lines 
\citep[e.g.,][] {thomas_05, bedregal_08, walcher_09, kobayashi_99, kuntschner_10}. 
They found that 
the  metallicity and [$\alpha$/Fe]  of stars in the core regions of 
early-type galaxies increase with the galactic mass.
The derived stellar metallicity is about a 1 solar in giant elliptical galaxies, 
considering the gradient of  strength of absorption lines
 \citep[]{arimoto_97, kobayashi_99, kuntschner_10}. 
However, the observations of absorption lines are limited to $\sim r_{\rm e}$, which corresponds 
to a half light radius. 
In addition, there may be systematic uncertainties in the assumption of the
age-population of stars, and in atomic physics.
For example, 
\citet{schiavon_07} found that optical spectra of some early-type galaxies
are better reproduced with a two-component age model, old and a small amount of
young population. Then, the derived metal abundances also changed from
a single component age model.
With X-ray observations, we can measure the abundances of O and Mg of X-ray
 emitting ISM  of the entire region of each galaxy.
The metals in the hot ISM in early-type galaxies are a mixture of those
from stars and recent SNe Ia and therefore, 
we can constrain stellar metallicity and their abundance pattern of 
the entire region of each galaxy.
The relatively simple atomic data for X-ray lines and temperature structure
of the ISM   reduce the systematic uncertainties in 
the abundance measurements.
Therefore, the X-ray measurements of metal abundances in the hot ISM 
and optical measurements of stellar metallicity are complementary.

In this paper, we performed uniform investigations of  the metal abundances in the ISM of 
17 early-type galaxies, 13 ellipticals and 4 S0's,  with {\it Suzaku}. 
This paper is structured as follows. In Section 2, we summarize observations of {\it Suzaku}. 
Sections 3 and 4 detail the data analysis and results. 
In Section 5, we investigated systematic differences between 
the v2.0.1 and v1.3.1 of the APEC plasma codes \citep[]{smith_01, foster_12} 
on the derived elemental abundances.
Section 6 gives a discussion of these results. 
Finally, in Section 7 we present our conclusion. 
Unless noted otherwise, we use the solar abundances in \citet{lodders_03}, the v2.0.1 of the 
APEC plasma codes, 
and the quoted errors are for a 90\% confidence interval for a single interesting parameter.

\section{TARGETS AND OBSERVATIONS}
\label{sec:target}

We analyzed archival data of 17 early-type galaxies observed with {\it Suzaku}.
The sample consists of 4 S0's and 13 ellipticals, whose characteristics and 
observational log  are summarized in Tables \ref{tbl:catalog} and \ref{tbl:log}, respectively. 
Our sample includes only luminous early-type galaxies  with $L_K>10^{11}L_\odot$.
The luminosity of $10^{11}L_\odot$ in K band close to
$L_*$, or  the characteristic luminosity of
the luminosity function by \citet{schechter_76} 
 for  clusters and groups galaxies by \citet{lin_04}.
Nine galaxies are located in cluster environments (Virgo and Fornax clusters), 
while the others are either in the field or in small groups.
NGC~1399 is the cD galaxy of the Fornax cluster, and 
NGC~4472 is the central galaxy of the south subgroup in the Virgo cluster. 
The temperature profiles of the ISM of NGC~1399, NGC~4472, NGC~4636, and
NGC~5846  increase with radius \citep[e.g.][]{matsushita_00, nagino_09}.
These galaxies are the central objects of larger potential 
structure and  have significantly higher ISM luminosities.
Hereafter, we denote these galaxies as $X_{E}$ galaxies.

We used all available data of the galaxies in the sample with XIS.
The XIS consists of three front-illuminated (FI: XIS0, XIS2 and XIS3) CCD cameras 
and one back-illuminated (BI: XIS1) CCD camera \citep{koyama_07}.
The XIS2 detector suffered catastrophic damage on  November 2006.
During observations of the sample galaxies,
the XIS was operated in normal clocking mode (8~s exposure per frame), 
with the standard 
5 $\times$ 5 and 3 $\times$ 3 editing mode.

\begin{table*}
\caption{\rm Galaxy sample in the {\it Suzaku} archive data. \label{tbl:catalog}}
\begin{center}
\begin{tabular}{ccccccccccc}
\tableline\tableline\noalign{\smallskip}
Galaxy   & \multicolumn{1}{c}{$\rm type$\tablenotemark{a}} & \multicolumn{1}{c}{$\log~L_{\rm K}$\tablenotemark{b}}& 
\multicolumn{1}{c}{$r_{\rm e}$\tablenotemark{c}}   &     & \multicolumn{1}{c}{$\sigma$\tablenotemark{d}} & \multicolumn{1}{c}{$N_{\rm H}$\tablenotemark{e}} & 
 \multicolumn{1}{c}{$z$\tablenotemark{f}} & \multicolumn{1}{c}{$D$\tablenotemark{g}} & \multicolumn{1}{c}{$\log~L_{\rm X}$} & Note \\ 
         &               & $L_{\odot}$             &  arcsec                   & kpc & km/sec                & 10$^{20}$cm$^{-2}$ &   &Mpc   & erg/sec     & \\ 
\noalign{\smallskip}\tableline\noalign{\smallskip}
NGC~720  & -5.0 &   11.05  & 36.07  & 4.26 & 240 & 1.54 & 0.005821&  20.3& 40.46\tablenotemark{h} & \\ 
NGC~1316 & -2.0 &11.56 & 80.75  & 9.61 & 250 & 1.89 & 0.005871&  16.9 & 40.63\tablenotemark{h} &  Fornax\\
NGC~1332 & -2.0 &11.02& 28.00  & 2.88 & 319 & 2.23 & 0.005084& 17.7 & 40.12\tablenotemark{h} &  \\
NGC~1399 & -5.0 &11.27 & 40.47  & 3.93 & 362 & 1.34 & 0.004753& 16.9 & 41.13\tablenotemark{h} & $X_{\rm E}$, Fornax\\
NGC~1404 & -5.0 &11.07& 23.83  & 3.15 & 245  & 1.36 & 0.006494&  16.9  & 40.81\tablenotemark{i} & Fornax\\
NGC~1553 & -2.0 &11.08 & 65.63  & 4.79 & 185    & 1.50 & 0.003602& 13.4 & 39.92\tablenotemark{i} & \\
NGC~2300 & -5.0 &11.28 & 31.41  & 4.05 & 252    & 5.27 & 0.006354& 31.0 & 41.90\tablenotemark{j} & \\
NGC~3923 & -5.0 &11.57 & 49.79  & 5.87 & 241 & 6.21 & 0.005801&  25.8 & 40.74\tablenotemark{h} & \\
NGC~4125 & -5.0 &11.36 & 58.50  & 5.38 & 165 & 1.84 & 0.004523& 24.2 & 40.67\tablenotemark{i} & \\
NGC~4382 & -2.0 &11.33 & 54.59  & 2.73 & 187 & 2.52 & 0.002432& 16.8 & 40.25\tablenotemark{h} & Virgo\\
NGC~4406 & -5.0 &11.35 & 104.02 & 1.77 & 250 & 2.62 & 0.000814& 16.8 & 41.64\tablenotemark{i} & Virgo\\
NGC~4472 & -5.0 &11.63 & 104.02 & 7.07 & 302 & 1.66 & 0.003326& 16.8 & 41.40\tablenotemark{h} & $X_{\rm E}$, Virgo\\
NGC~4552 & -5.0 &11.10 & 29.31  & 0.67 & 262 & 2.57 & 0.001134& 16.8 & 40.62\tablenotemark{h} & Virgo\\
NGC~4636 & -5.0 &  11.23 & 88.54  & 5.67 & 165 & 1.81 & 0.003129&  17.0 & 41.46\tablenotemark{h} & $X_{\rm E}$, Virgo\\
NGC~4649 & -5.0 &  11.50 & 68.73  & 5.22 & 343 & 2.20 & 0.003726&  16.8  & 41.00\tablenotemark{h} & Virgo\\
NGC~4697 & -5.0 &  11.53 & 71.97  & 6.05 & 165 & 2.12 & 0.004140& 23.3 & 40.23\tablenotemark{i} & \\
NGC~5846 & -5.0 & 11.48 & 62.68  & 7.27 & 251 & 4.26 & 0.005717& 28.5 & 41.72\tablenotemark{h} & $X_{\rm E}$ \\
\noalign{\smallskip}\tableline\noalign{\smallskip}
\end{tabular}
\tablenotetext{a}{Morphological type code from \citet{tully_88}.}
\tablenotetext{b}{Calculated K-band luminosity from the Two Micron All Sky Survey (2MASS). The effect of Galactic extinction was corrected using the NASA/IPAC Extragalactic Database (NED).}
\tablenotetext{c}{Effective radius from RC3 Catalog \citep{vaucouleurs_91}.}
\tablenotetext{d}{Central stellar velocity dispersion from \citet{prugniel_96}.}
\tablenotetext{e}{Column density of the Galactic absorption from \citet{dickey_90}.}
\tablenotetext{f}{Redshift from Nasa Extragalactic Database.}
\tablenotetext{g}{Distance from \cite{tully_88}.}
\tablenotetext{h}{X-ray luminosity of the thermal emission in the range of 0.3--2.0 keV within 4$r_{\rm e}$ from \citet{nagino_09}.}
\tablenotetext{i}{X-ray luminosity of the thermal emission in the range of 0.2--2.0 keV within 4$r_{\rm e}$ from \citet{Matsushita_01}.} 
\tablenotetext{j}{X-ray luminosity of the thermal emission in the range of 0.4--2.0 keV within 25$'$ from \citet{davis_96}.}
\end{center}
\end{table*}

\begin{table*}
\caption{\rm Observational log.\label{tbl:log}}
\begin{center}
\begin{tabular}{ccccccccccc}
\tableline\tableline\noalign{\smallskip}
Galaxy   & \multicolumn{1}{c}{$\rm Obs ID$\tablenotemark{a}} & \multicolumn{1}{c}{$\rm Obs date$\tablenotemark{b}} & \multicolumn{1}{c}{$\rm exposure$\tablenotemark{c}}    \\ 
         &        &          &     (ks)          \\
\noalign{\smallskip}\tableline\noalign{\smallskip}
NGC~720  &800009010 & 2005-12-30 &177.2  \\
NGC~1316 &801015010& 2006-12-22 &48.7   \\
NGC~1332 &805095010  	 & 2011-01-20  & 101.4   \\
NGC~1399 \& NGC1404 &100020010 &2005-09-13 & 76.1   \\
NGC~1553 &802050010 & 2007-11-25 &98.7   \\
NGC~2300 &804030010 & 2010-02-08  & 37.1   \\
         &804030020 & 2010-02-10 & 52.6   \\
NGC~3923 &801054010 & 2006-06-13 &115.8   \\
NGC~4125 &804047010 & 2009-09-29 &104.1   \\
NGC~4382 &803005010 & 2008-06-21  &99.1   \\
NGC~4406 &803043010 & 2009-06-19  &101.8  \\  
NGC~4472 &801064010 & 2006-12-03  &121.0  \\
NGC~4552 &701037010 & 2006-12-03  &20.4   \\
NGC~4636 &800018010 & 2005-12-06  &79.2   \\
NGC~4649 &801065010 & 2006-12-29  &224.0  \\
NGC~4697 &805041010 & 2011-01-14  &102.3  \\
NGC~5846 &803042010 & 2008-07-28  &155.9  \\
\noalign{\smallskip}\tableline\noalign{\smallskip}
\tablenotetext{a}{Observation number of the {\it Suzaku} data.}
\tablenotetext{b}{Observation start date.}
\tablenotetext{c}{Exposure time after data screening (ksec).}
\end{tabular}
\end{center}
\end{table*}

\section{DATA REDUCTION AND CONSTRUCTION OF SPECTRA}
\label{sec:data}
                        
We processed the XIS data using the \verb+xispi+ and \verb+makepi+ ftool tasks 
and CALDB files of version 2012-07-03. 
The XIS data  with the Earth 
elevation angles less than $5^{\circ}$ or the Day-Earth elevation angles less
than $20^{\circ}$ were excluded.
We also discarded data with 
time since south Atlantic anomaly passage of less than 436 sec. 
We created a 0.3--5 keV light curve for each sensor, with 256 s binning. 
All galaxies have no periods of anomalous event rates 
higher or less than $\pm3 \sigma$ from the mean. 
After this screening, 
the remaining good exposures were listed in Table \ref{tbl:log}.  
Event screening with cut-off rigidity was not performed.
The spectral analysis was performed with HEAsoft version 6.12 and XSPEC 12.7.

\begin{figure*}
\centering
\includegraphics[width=0.8\textwidth,angle=0,clip]{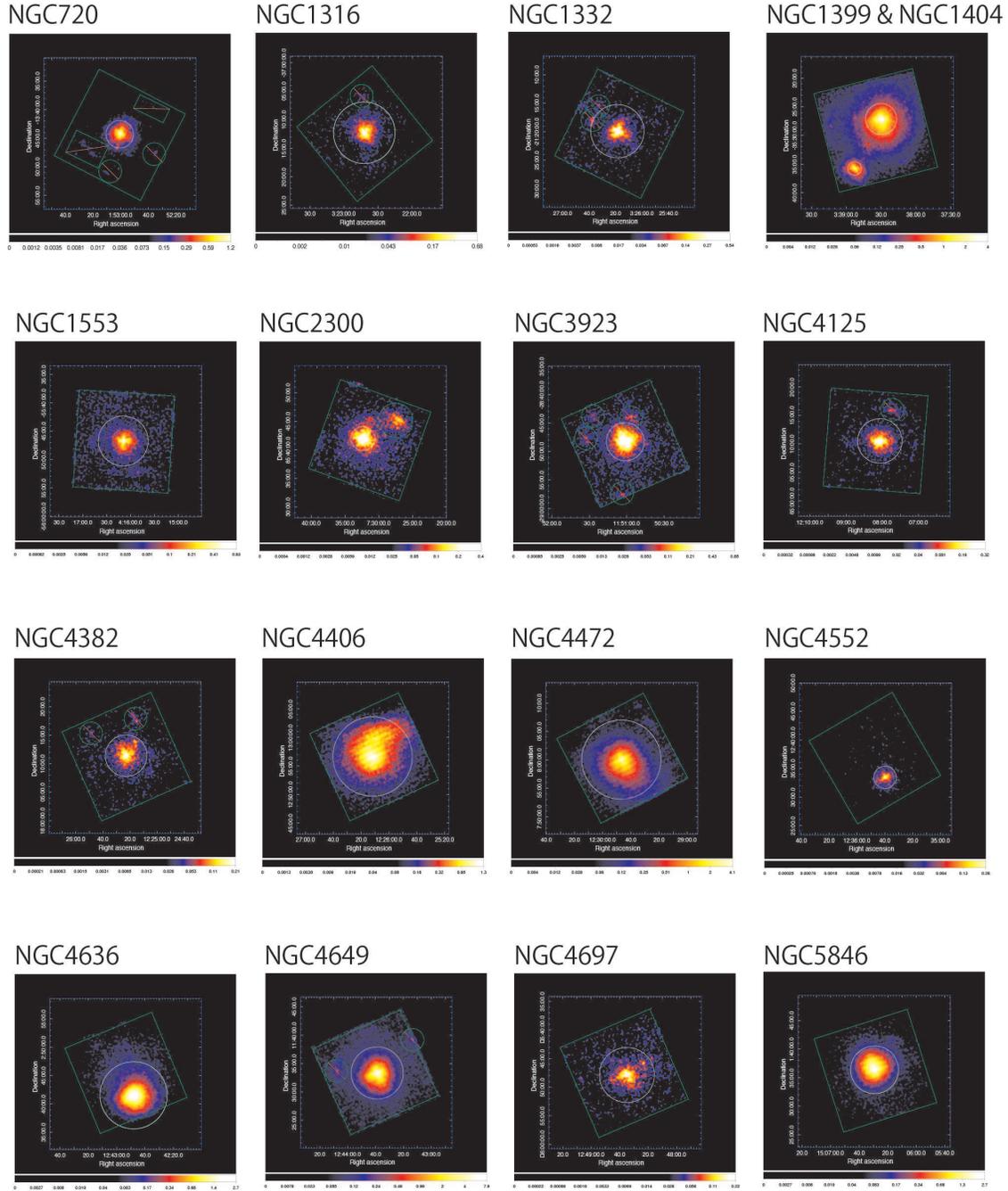}
\caption{
The 0.4--5 keV XIS1 images of our samples, smoothed with a $\sigma$=7\farcs5 Gaussian profile. 
Background components were not subtracted, and vignetting was not corrected in these images. 
The on-source regions are  circular regions with a 4 $r{_e}$ radius 
 shown in  white circles. 
The green box and green circles represent
 the entire XIS field of view and the regions excluded from the
 analysis, respectively.
}
\label{img:img}
\end{figure*}

We accumulated  on-source spectra for each galaxy
within 4 $r_{\rm e}$ centered on each galaxy.
We used a accumulation radius of 3 $r_{\rm e}$   for NGC~1404,
which is located near the cD galaxy, NGC 1399, of the Fornax cluster.
To study the background emission,
we also accumulated spectra over the entire XIS field of view
excluding the on-source region (hereafter the background regions).
The 4 $r_{\rm e}$ regions, which include most of stars in individual  galaxies,
are suitable for Suzaku analysis considering the point spread function.
A larger accumulating region may suffer from a larger systematic uncertainties in the
background emission including the cluster and group emission.
Regions around luminous point sources were excluded from the analysis.
The region around  NGC~1404 was also excluded in the analysis of  NGC~1399.
We adopted an annular region of  3 $r_{\rm e}$--6 $r_{\rm
e}$ from the center of NGC~1404 as the background region for NGC~1404.
In Figure \ref{img:img}, we show  XIS images with the  on-source
 and the excluded regions for the analysis.
Because the ISM emits little photons above 5 keV, 
to improve the photon statistics,
we included corner regions  illuminated by calibration sources,
which have an emission peak  at 5.9 keV.

To subtract the non-X-ray background (NXB), we employed 
the Dark-Earth database using the \verb+xisnxbgen+ ftool task \citep{tawa_08}.
Redistribution matrix file (RMF) file  were calculated using 
the  \verb+xisrmfgen+ ftool task \citep{ishisaki_07}, version 2011-07-02. 
The low energy transmission of the XIS optical blocking filters
(OBF)  has been decreased due to contamination on the filters \citep{koyama_07}. 
The contaminant thickness has been evolved with time,
and  depends on the detector position.
We used the  ancillary response files (ARFs) generator, 
\verb+xissimarfgen+ ftool task \citep{ishisaki_07}, version 2010-11-05, 
which  includes transmission through the contaminant.
Here, we used  the observed XIS1 image of each galaxy 
 for the emission components  from the galaxy.
For  background components, we calculated ARF files in the same way 
but assuming a uniform sky emission.
However, for the data of two X-ray luminous galaxies,  NGC~4472 and NGC~4649
observed in December 2006 
and all the data taken since 2008,
there are still discrepancies between the data and model below 0.6 keV
even using ARFs including the contaminant effect.
Therefore, for these data, we used ARF files without including the
effect of the contaminant.
Instead,  we  multiplied 
``varabs'' model,  a photoelectric absorption model
with variable abundances, by models used for spectral fitting.
We summarized the treatment of contaminant and setting parameters of 
the ``varabs'' model in Appendix A.

\section{Analysis and Results}
\label{sec:analysis}

\subsection{Spectral fits with Single- and Two-Temperature models for the ISM}
\label{ssec:1t2t}

\subsubsection{Estimation of the background spectra}
\label{sssec:bgd}

In order to estimate background components, we first fitted the 
spectra of the background region. 
We used the following model for the emission from the background region: 
phabs $\times$ (power-law$_{\rm CXB}$ + apec$_{\rm ETE}$ + apec$_{\rm MWH}$) + 
apec$_{\rm LHB}$. 
In this model, ``phabs'' represents the Galactic 
absorption in the direction of each galaxy. 
After subtracting the NXB component,
the emission of background regions consists of
the  cosmic X-ray background (CXB) 
and the emission from our Galaxy.
We assumed a power-law model, ``power-law$_{\rm CXB}$''
 for the CXB component with a slope of $\Gamma=1.4$
\citep{kushino_02}, and  two  APEC model components,
 ``apec$_{\rm LHB}$'' and ``apec$_{\rm MWH}$'',
 for  the  Galactic emission.
The  ``apec$_{\rm LHB}$'' component represents 
the sum of the solar wind charge exchange
(SWCX) and local hot bubble (LHB), and
the   ``apec$_{\rm MWH}$'' corresponds to the emission from  Milky Way halo (MWH)
 \citep{yoshino_09}. 
The metal abundances and redshift of
  ``apec$_{\rm MWH}$'' and ``apec$_{\rm LHB}$'' components were fixed at
  the solar value and zero, respectively.
Though the temperatures of ``apec$_{\rm LHB}$'' and ``apec$_{\rm MWH}$'' models are basically 
set to free, 
for several galaxies with large error bars in temperatures of the Galactic components, 
we fixed theses temperatures at the values, which have $\chi^2$ minimum within a range of 
0.05--0.3 keV and 0.3--0.6 keV for LHB and MWH, respectively.

The ISM emission may extend beyond each source region and
galaxies located in clusters and groups are surrounded with intracluster
medium (ICM).
Therefore, we added an APEC component,  ``apec$_{\rm ETE}$'', 
for these extra thermal emission (ETE).
The metal abundances of ``apec$_{\rm ETE}$'' components were free, 
but for NGC~1316, NGC~4406, and NGC4552, were fixed to 0.3 solar, 
because they are surrounded by the ICM \citep{tashiro_09, shibata_01, machacek_06}. 
The redshift of  ``apec$_{\rm ETE}$'' component was fixed at
that of each galaxy shown in  Table \ref{tbl:catalog}.
To fit the spectra of the background region,
we need a two-temperature ``apec$_{\rm ETE}$'' model for
the ETE emission around  NGC 4406, NGC 4472,
and NGC 5846, and a two-temperature ``vapec$_{\rm ETE}$'' model for NGC 1399,
and a single-temperature``vapec$_{\rm ETE}$'' model for NGC 4636. 
The ``vapec$_{\rm ETE}$'' model are variable abundance model adopting APEC plasma code 
\citep{smith_01}.
We summarized the derived parameters in Table \ref{tbl:bgdonly1} and \ref{tbl:bgdonly2} 
in Appendix \ref{a:analysis}.

\subsubsection{Simultaneously fit of the source and background spectra}
\label{sssec:simfit}

\begin{figure*}[tbp]
 \begin{center}
\includegraphics[width=0.8\textwidth,angle=0,clip]{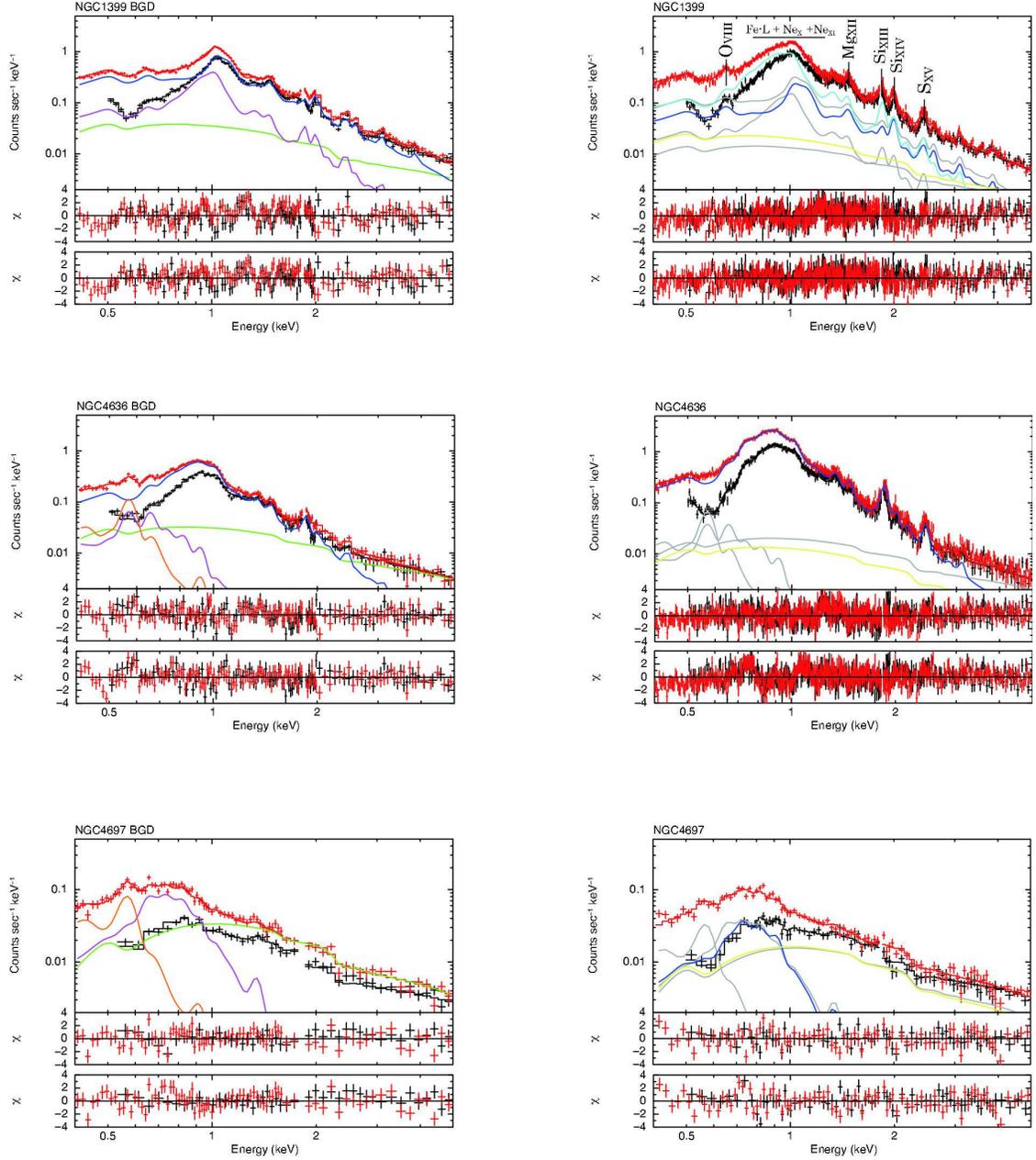}
 \end{center}
 \caption{NXB-subtracted XIS0 (black) and XIS1 (red) spectra of each
 background region (left column),  and those of NGC~1399 (1.05 keV), NGC4636 (0.80 keV), and
 NGC~4697 (0.25 keV) within the 4 $r_{\rm e}$ region (right column) 
in order from top to bottom, shown without removing the instrumental responses. The black and red lines show the best-fit model for the XIS0 and XIS1, respectively. For simplicity, only the model components for the XIS1 spectra are shown.  
The blue and light blue lines show  the ISM components, 
yellow corresponds to the emission from discrete point sources,
the magenta and orange lines show contributions from the Galactic  emission 
by ``apec$_{\rm MWH}$'' and ``apec$_{\rm LHB}$'', respectively,
and the  green lines for the CXB components.
Panels under spectra show residuals of fit utilizing AtomDB v2.0.1 and v1.3.1 
from the upper and the lower panels, respectively. 
}
\label{img:spec}
\end{figure*}

The emission from each on-source region consists of
 the source emission from individual galaxies and the background emission.
We assumed a following model as the source emission; 
phabs$\times$ (vapec$_{\rm ISM}$ + power-law$_{\rm PS}$). 
Here, the ``phabs'' model corresponds to the photoelectric-absorption,
whose  column density was  fixed to the Galactic value 
in the direction of  each galaxy, shown  in Table 1. 
The ``vapec$_{\rm ISM}$'' model means thermal emission from the ISM.
As shown in Figure \ref{img:spec}, several emission lines seen
 around 0.5--0.6 keV, 0.6--0.7 keV, 
$\sim$ 1.3 keV, and $\sim$ 1.8 keV are identified with K$\alpha$ lines of 
 \ion{O}{7},  \ion{O}{8}, \ion{Mg}{11},  \ion{Si}{13}, respectively.
The emission bump around 0.7--1 keV mostly corresponds to
the Fe-L complex, with smaller contributions by K-lines from \ion{Ne}{9} and  \ion{Ne}{10}
and the Ni-L complex.
 We divided the  metals into 
five groups as O, Ne, (Mg \& Al), (Si, S, Ar, Ca), and (Fe \& Ni), 
based on the metal synthesis mechanism of SNe, and allowed 
each group to vary. 
For luminous galaxies, NGC~1399, NGC~4406, NGC~4472, NGC~4649, and NGC~5846, 
we divided Si group as Si and (S, Ar, Ca). 
We set an upper limit of 5 solar for metal abundances.
The abundances of He, C, and N were fixed  to the solar value.
The ``power-law$_{\rm PS}$'' component represents the integrated discrete sources, 
with its photon index being fixed at 1.6 \citep[e.g.,][]{xu_05, randall_06}.

We used the same model in Section \ref{sssec:bgd} for the emission from the background region. 
The metal abundance of the ``apec$_{\rm ETE}$'' was fixed at the best-fit value, 
which is summarized in Table \ref{tbl:bgdonly1} (Appendix \ref{a:analysis}), 
from fitting for the background regions. 
Only for NGC~1399 and NGC~4636, each metal abundances of ETE emission, which are represented 
by ``vapec'' models,
are set to free, but limited within errors derived in Section \ref{sssec:bgd}, 
as shown in Table \ref{tbl:bgdonly2} (Appendix \ref{a:analysis}). 

We simultaneously fitted the spectra of the on-source and background regions
to take into account background emission accurately. 
The background components (CXB, LHB, and MWH) were assumed to have the same surface brightness 
between the two regions.
Except for NGC~1316, NGC~4406, and NGC~4552, the normalizations of the ETE components
in source regions have been set to zero.
We assumed the same surface brightness for the ETE components in the source and
the background regions of these three galaxies, which are surrounded by the ICM emission.
For NGC~1404, we subtracted the spectra of the annular region of
3--6 $r_{\rm e}$ as the background.
We used  energy ranges of 0.4--5.0 keV  and 0.5--5.0 keV of
the BI and  FI detectors, respectively.
Because there are known calibration issues of the Si edges in all XIS CCDs, 
an energy range of 1.84--1.86~keV is ignored \citep{koyama_07} in {\it Suzaku} spectra.

\begin{figure*}[tbp]
 \begin{center}
\includegraphics[width=0.5\textwidth,angle=0,clip]{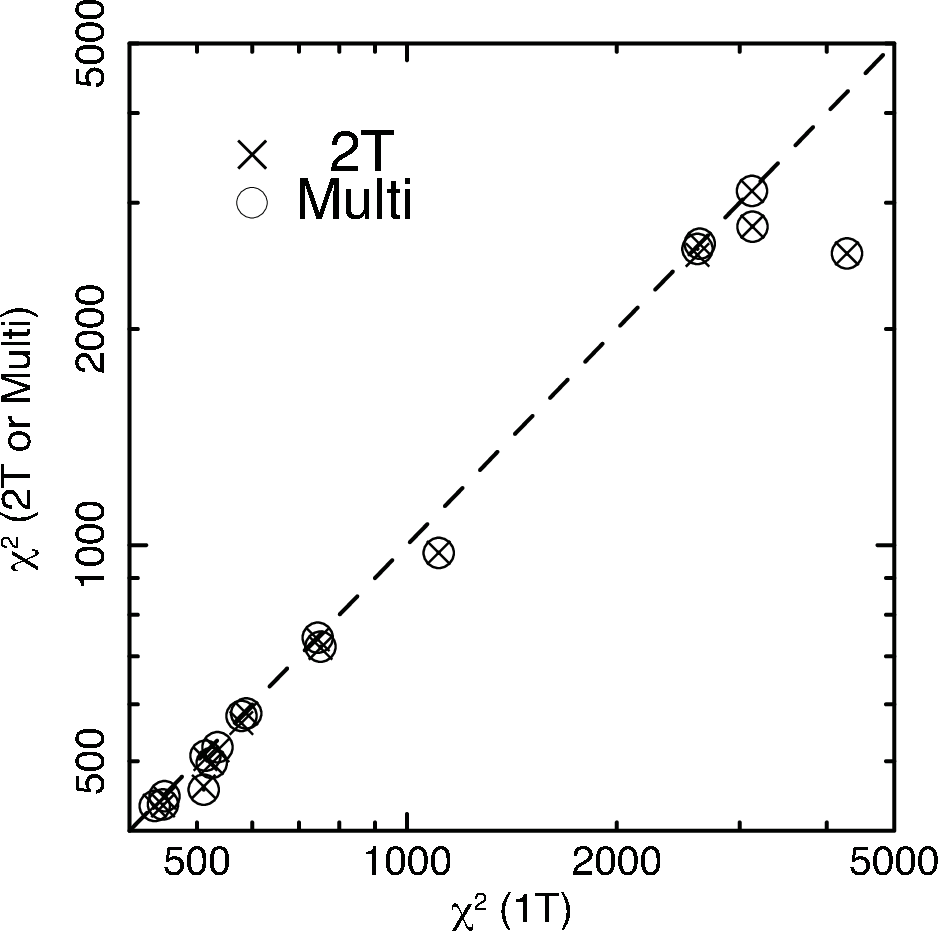}
 \end{center}
 \caption{Comparison of  ${\chi^2}$ between the 1T and 2T or multi-T models. The dashed lines indicate the equal value between the two.
}
\label{img:chi2_1t_vs_2tmulti}
\end{figure*}

We used both one-temperature (1T) and two-temperature (2T) plasma models
for the ISM emission to fit the spectra.
For the 2T model, metal abundances of each element  
for the two temperature components were assumed to have a same value.
The representative spectra fitted in this way are shown in Figure \ref{img:spec}.
The spectra of the other galaxies are shown in Appendix \ref{a:analysis}.
The 1T or 2T models for the ISM represent the spectra fairy well, expect
for residual structures around 1.2~keV.
Table \ref{tbl:gal} summarizes  the derived ISM temperatures  and 
$\chi^2$ for  the spectral fits.
With the 1T model fits, 
the temperatures of ISM of the sample galaxies  range from 0.25 to 1.1 keV. 
As shown in Figure \ref{img:chi2_1t_vs_2tmulti}, the resultant  $\chi^2$  fitted with
the single- and two-temperature models for the ISM are almost the same,
except for several galaxies.
For ten galaxies, NGC~1316, NGC~1399, NGC~1404, NGC~1553, NGC~2300, NGC~3923, NGC~4125, NGC~4406, 
NGC~4472, and NGC~5846, we adopted the results of 2T model fits,
since their F-test probabilities for adding the second temperature component are 
lower than 1\%. 
Among them, NGC~1399, NGC~4472, and NGC~5846 are $X_{\rm E}$ galaxies, with
 positive temperature gradients \citep[][]{nagino_09}.
NGC~4406 is an X-ray luminous galaxies with very complicated X-ray
emission caused by ram-pressure stripping with the ICM in the Virgo
cluster \citep[e.g.,][]{matsushita_00, Matsushita_01}.
The metal abundances derived from 1T and 2T model are summarised in Section \ref{ssec:abund_1t2t}. 
The reduced $\chi^2$ values with the 1T or 2T model fits
 are about 1.05--1.37 except for NGC 5846 whose reduced $\chi^2$ is 1.73.
Considering only  statistical errors, most of the fits are not acceptable.
The large $\chi^2$ mainly caused by the residual structures around 1.2 keV.
We  discuss these residuals and possible systematic uncertainties
in the Fe-L modeling  in Section \ref{ssec:0.8kev} and \ref{ssec:1.2kev}

Although the intensities of ``power-law$_{\rm PS}$'' components have low impact to metal abundance 
measurements, we have checked these luminosities. 
The derived luminosities of ``power-law$_{\rm PS}$'' components 
are compared to those derived 
from {\it Chandra} observations by \citet{boroson_11}. 
Overlapping their samples, the luminosities of ``power-law$_{\rm PS}$'' components of 
NGC~1316, NGC~4125, NGC~4382, and NGC~4649, 
are good agreement with sum luminosities of 
low mass X-ray binaries and active galactic nuclei in \citet{boroson_11}.

\begin{table*}
\caption{
\rm The ISM temperatures and $\chi^2$ derived from the 1T, 2T and multi-T model fits
using APEC plasma code, v2.0.1. The flux ratios of high and low temperature ISM for galaxies whose best fit is 2T model. 
 \label{tbl:gal}}
\begin{center}
\begin{tabular}{ccccccccccc}
\tableline\tableline\noalign{\smallskip}
Galaxy   & kT (1T) & ${\chi^2}$/d.o.f. &  kT (2T) &            &  flux ratio      &   ${\chi^2}$/d.o.f. & ${\chi^2}$/d.o.f.  \\
         & (keV)   &         (1T)      &   (keV)  &   (keV)  & high$_{\rm T}$/low$_{\rm T}$   &    (2T)          & (Multi-T)      \\
\noalign{\smallskip}\tableline\noalign{\smallskip}
NGC~720  & 0.61$^{+0.01}_{-0.01}$ & 745/545   &     0.62$^{+0.07}_{-0.02}$ &    $<$0.41   & \nodata   & 739/543 &   743/547   \\
NGC~1316 & 0.65$^{+0.02}_{-0.02}$ & 511/414   & 0.93$^{+0.04}_{-0.04}$     &  0.45$^{+0.05}_{-0.05}$  & 1.09  &  461/412   &  411/411   \\
NGC~1332 & 0.60$^{+0.02}_{-0.02}$ & 449/411   &   0.86$^{+0.50}_{-0.10}$   &  0.53$^{+0.09}_{-0.43}$ & \nodata   & 445/409      &    447/412    \\
NGC~1399 & 1.05$^{+0.01}_{-0.01}$ & 3129/2023 &   1.74$^{+0.22}_{-0.05}$   & 1.01$^{+0.01}_{-0.01}$ & 0.39  &2778/2021   &    2780/2028    \\
NGC~1404 & 0.66$^{+0.01}_{-0.01}$ & 447/407   &  0.72$^{+0.08}_{-0.06}$    & 0.43$^{+0.03}_{-0.02}$  &  3.64   & 434/405       &   435/406    \\
NGC~1553 & 0.57$^{+0.02}_{-0.02}$ & 535/415   &  0.64$^{+0.09}_{-0.09}$     & 0.31$^{+0.08}_{-0.08}$& 0.85  & 516/412         &  523/415    \\
NGC~2300 & 0.74$^{+0.02}_{-0.02}$ & 579/442   & 4.06$^{+1.30}_{-2.21}$    & 0.73$^{+0.02}_{-0.02}$  & 0.33    &566/440        &  578/443  \\
NGC~3923 &  0.60$^{+0.01}_{-0.01}$&  752/547  &   0.41$^{+0.04}_{-0.03}$  &  0.70$^{+0.04}_{-0.08}$&  0.94   &719/545  & 722/547  \\
NGC~4125 & 0.47$^{+0.03}_{-0.03}$ & 525/408   &   0.59$^{+0.17}_{-0.08}$&  0.32$^{+0.10}_{-0.07}$& 0.91  & 497/406           &  497/410   \\
NGC~4382 & 0.33$^{+0.03}_{-0.02}$ & 435/413   &  $<$0.39               &  $<$0.26          & \nodata   & 433/411           &  433/415  \\
NGC~4406 & 0.84$^{+0.01}_{-0.01}$ &  2612/2069& 1.29$^{+0.22}_{-0.07}$ & 0.81$^{+0.01}_{-0.01}$& 0.34    &  2537/2067      &   2586/2070   \\
NGC~4472 & 1.04$^{+0.01}_{-0.01}$ & 3288/2160 & 1.65$^{+0.08}_{-0.09}$  & 1.00$^{+0.01}_{-0.01}$& 0.43   & 2549/2158         &  2549/2160  \\
NGC~4552 & 0.65$^{+0.04}_{-0.03}$ & 588/485   & 0.75$^{+0.13}_{-0.11}$ & 0.38$^{+0.22}_{-0.11}$ & \nodata   & 583/483             &  583/487  \\
NGC~4636 & 0.80$^{+0.01}_{-0.01}$ & 3124/2298 &  0.61$^{+0.03}_{-0.07}$ &  0.85$^{+0.02}_{-0.03}$   & \nodata   &3115/2301              & 3114/2303  \\
NGC~4649 & 0.86$^{+0.01}_{-0.01}$ & 2630/2057 &  0.85$^{+0.01}_{-0.01}$ &   $<$0.12     & \nodata   & 2623/2055   & 2630/2057  \\
NGC~4697 & 0.25$^{+0.05}_{-0.03}$ & 514/412   &     $<$0.76           & 0.19$^{+0.04}_{-0.05}$    & \nodata   &  503/410           &  509/414   \\
NGC~5846 & 0.76$^{+0.01}_{-0.01}$ &  1110/567 & 0.88$^{+0.01}_{-0.01}$ & 0.58$^{+0.03}_{-0.03}$ &  1.69    &  976/565      &  976/566  \\
\noalign{\smallskip}\tableline\noalign{\smallskip}
\end{tabular}
\end{center}
\end{table*}

\begin{figure*}[tbp]
 \begin{center}
\includegraphics[width=\textwidth,angle=0,clip]{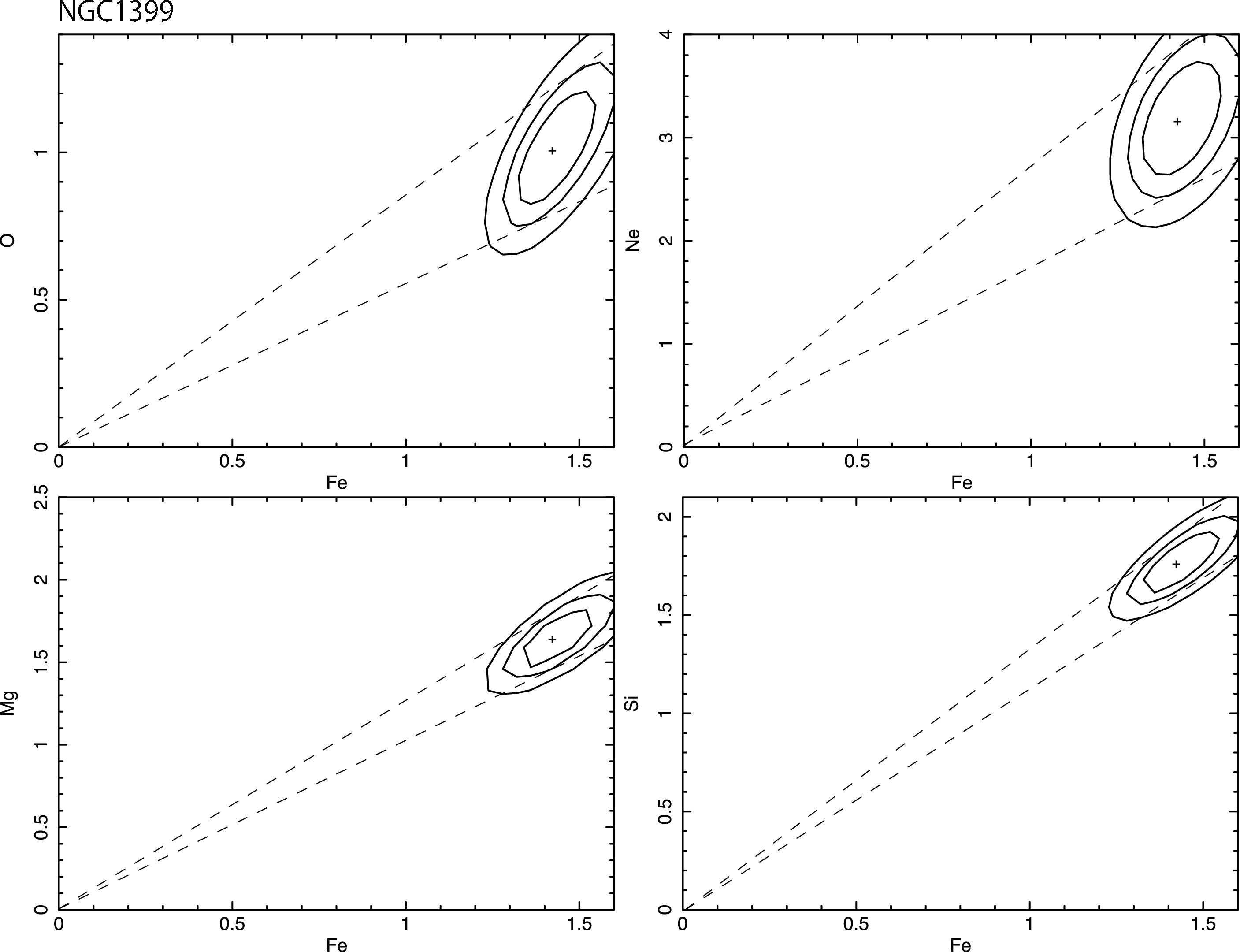}
 \end{center}
 \caption{The confidence contours of the abundances of  O, Ne, Mg, Si vs. 
Fe  in solar units for  NGC~1399 with the 2T model fits.
The three contours represent the 68\%, 90\%, and 
99\% confidence ranges from inner to outer, respectively. The dashed lines show the 
90\%-confidence level of these metal abundances relative to Fe.
}
\label{img:contour}
\end{figure*}

In order to examine the abundance ratios rather than their absolute vales, we calculated the confidence contours 
between the abundance of various metals (O, Ne, Mg, and Si) to that of Fe. 
The results are shown in Figure \ref{img:contour}, where we also indicate 90\%-confidence abundance 
of these metals relative to Fe with black dashed line. The elongated shape of the confidence contours indicates 
that the relative values were determined more accurately than the absolute values. 



The resultant temperatures in the background components
are consistent to those derived in Section \ref{sssec:bgd} (Table \ref{tbl:bgdonly1}). 
The derived  of the  Galactic components, MWH and LHB (and SWCX), are 
about 0.2--0.3 keV and 0.1 keV, respectively.
These values are consistent with those derived for  
other sky regions without luminous sources observed with {\it Suzaku} \citep[e.g.,][]{yoshino_09}. 

\subsection{Spectral fits with multi-temperature model for the ISM}
\label{ssec:multi}



The temperature gradients of the ISM within 4 $r_e$ in 
early-type galaxies have been reported with 
{\it Chandra} and {\it XMM-Newton} satellites \citep[e.g.,][]{nagino_09, ji_09}.
 X-ray luminous galaxies tend to have increasing temperature profiles, while others show
constant or decreasing profile.
We used a multi-temperature model (hereafter multi-T model) 
for the ISM and refitted the spectra
in the same way as in Section \ref{ssec:1t2t}.
For galaxies which need the 2T model for the ISM, we fitted the spectra
with a five-temperature model, and for the other galaxies, we used
a three-temperature model.
The difference of  temperatures of  the
two neighboring components was fixed at 0.2 keV,
and the temperature of the central temperature component  was fixed to be
the best-fit temperature derived from the 1T model fit or the intermediate value 
between the best-fit two temperatures derived from the 2T model fit. 
The multi-T model fit gave almost the same $\chi^2$ 
(Figure \ref{img:chi2_1t_vs_2tmulti})
and show similar residual structures with the
1T or 2T model fits.

\begin{table*}
\caption{
\rm The O, Ne, Mg, Si, S and Fe abundances in the ISM 
 derived from the multi-T model fits using APEC plasma code, v2.0.1.
 \label{tbl:metal_multi}}
\begin{center}
\begin{tabular}{ccccccccccc}
\tableline\tableline\noalign{\smallskip}
Galaxy   & O & Ne & Mg  & Si &  S & Fe  \\
         & (solar) & (solar)  & (solar)   & (solar)  & (solar) & (solar) \\ 
\noalign{\smallskip}\tableline\noalign{\smallskip}
NGC~720  & 1.09$^{+0.38}_{-0.26}$ & 1.82$^{+0.55}_{-0.38}$ & 1.16$^{+0.37}_{-0.27}$ & 1.19$^{+0.43}_{-0.30}$ & =Si  &  1.05$^{+0.28}_{-0.19}$  \\
NGC~1316 & 0.92$^{+0.54}_{-0.31}$ & 1.43$^{+0.98}_{-0.60}$ & 0.89$^{+0.65}_{-0.35}$ & 0.49$^{+0.43}_{-0.26}$   & =Si           & 0.79$^{+0.44}_{-0.24}$ \\
NGC~1332 & 3.19$^{+1.81}_{-1.62}$ & 5.00$^{+0}_{-2.25}$    & 3.76$^{+1.24}_{-1.74}$ & 1.92$^{+1.52}_{-1.16}$    & =Si       & 2.52$^{+1.15}_{-1.08}$  \\
NGC~1399 & 1.01$^{+0.17}_{-0.19}$ & 3.16$^{+0.41}_{-0.41}$ & 1.64$^{+0.16}_{-0.07}$ & 1.76$^{+0.26}_{-0.11}$ & 1.30$^{+0.28}_{-0.15}$   &1.43$^{+0.16}_{-0.08}$  \\
NGC~1404 & 1.20$^{+0.58}_{-0.36}$ & 1.67$^{+0.83}_{-0.58}$ & 0.89$^{+0.44}_{-0.28}$ & 1.52$^{+0.81}_{-0.51}$   & =Si    & 1.27$^{+0.47}_{-0.28}$  \\
NGC~1553 & 5.00$^{+0}_{-3.23}$    & 4.80$^{+0.20}_{-3.48}$ & 3.23$^{+1.77}_{-2.30}$ & 1 (fix)        & =Si         & 4.77$^{+0.23}_{-3.02}$  \\
NGC~2300 & 1.02$^{+0.50}_{-0.44}$ & 2.20$^{+0.81}_{-0.76}$ & 1.04$^{+0.45}_{-0.27}$ & 0.80$^{+0.35}_{-0.22}$   & =Si     & 0.60$^{+0.21}_{-0.13}$ \\
NGC~3923 & 1.21$^{+0.94}_{-0.42}$ & 1.80$^{+1.42}_{-0.76}$ & 1.46$^{+1.14}_{-0.48}$ & 1.11$^{+0.94}_{-0.51}$ & =Si      & 1.26$^{+0.87}_{-0.37}$  \\
NGC~4125 & 0.49$^{+0.28}_{-0.17}$ & 0.82$^{+0.45}_{-0.27}$ & 0.50$^{+0.31}_{-0.19}$ & 0.44$^{+0.48}_{-0.33}$ & =Si      & 0.60$^{+0.25}_{-0.14}$  \\
NGC~4382 & 0.83$^{+0.50}_{-0.28}$ & 1.28$^{+0.79}_{-0.44}$ & 1.12$^{+0.80}_{-0.46}$ & 1 (fix)         & =Si          & 1.29$^{+0.75}_{-0.40}$  \\
NGC~4406 & 0.80$^{+0.19}_{-0.17}$ & 2.28$^{+0.38}_{-0.36}$ & 0.79$^{+0.09}_{-0.10}$ & 0.53$^{+0.07}_{-0.03}$ & 0.79$^{+0.07}_{-0.09}$ &  0.72$^{+0.07}_{-0.06}$  \\
NGC~4472 & 0.86$^{+0.12}_{-0.10}$ & 2.07$^{+0.20}_{-0.11}$ & 0.88$^{+0.08}_{-0.07}$ & 0.96$^{+0.06}_{-0.05}$ & 0.98$^{+0.07}_{-0.07}$  &   0.86$^{+0.05}_{-0.04}$  \\
NGC~4552 & 1.08$^{+1.04}_{-0.56}$ & 0.85$^{+1.30}_{-0.85}$ & 0.95$^{+0.78}_{-0.45}$ & 1 (fix)      & =Si            & 0.83$^{+0.50}_{-0.25}$ \\
NGC~4636 & 0.73$^{+0.07}_{-0.07}$ & 1.97$^{+0.15}_{-0.15}$ & 0.88$^{+0.06}_{-0.05}$ & 0.86$^{+0.05}_{-0.05}$  & =Si   & 0.80$^{+0.04}_{-0.03}$  \\
NGC~4649 & 0.87$^{+0.12}_{-0.11}$ & 1.71$^{+0.16}_{-0.15}$ & 0.80$^{+0.08}_{-0.07}$ & 0.75$^{+0.06}_{-0.05}$ & 0.85$^{+0.09}_{-0.08}$ &  0.69$^{+0.04}_{-0.04}$  \\
NGC~4697 & 0.62$^{+0.82}_{-0.29}$ & 1.95$^{+1.98}_{-0.85}$ & 1 (fix)                & 1 (fix)      & =Si         & 1.46$^{+1.58}_{-0.61}$  \\
NGC~5846 & 1.03$^{+0.15}_{-0.12}$ & 2.14$^{+0.24}_{-0.20}$ & 1.02$^{+0.10}_{-0.09}$ & 0.98$^{+0.09}_{-0.08}$ & 1.16$^{+0.07}_{-0.13}$  &  0.84$^{+0.06}_{-0.05}$  \\
\noalign{\smallskip}\tableline\noalign{\smallskip}
\end{tabular}
\end{center}
\end{table*}

\begin{figure*}[tbp]
 \begin{center}
\includegraphics[width=\textwidth,angle=0,clip]{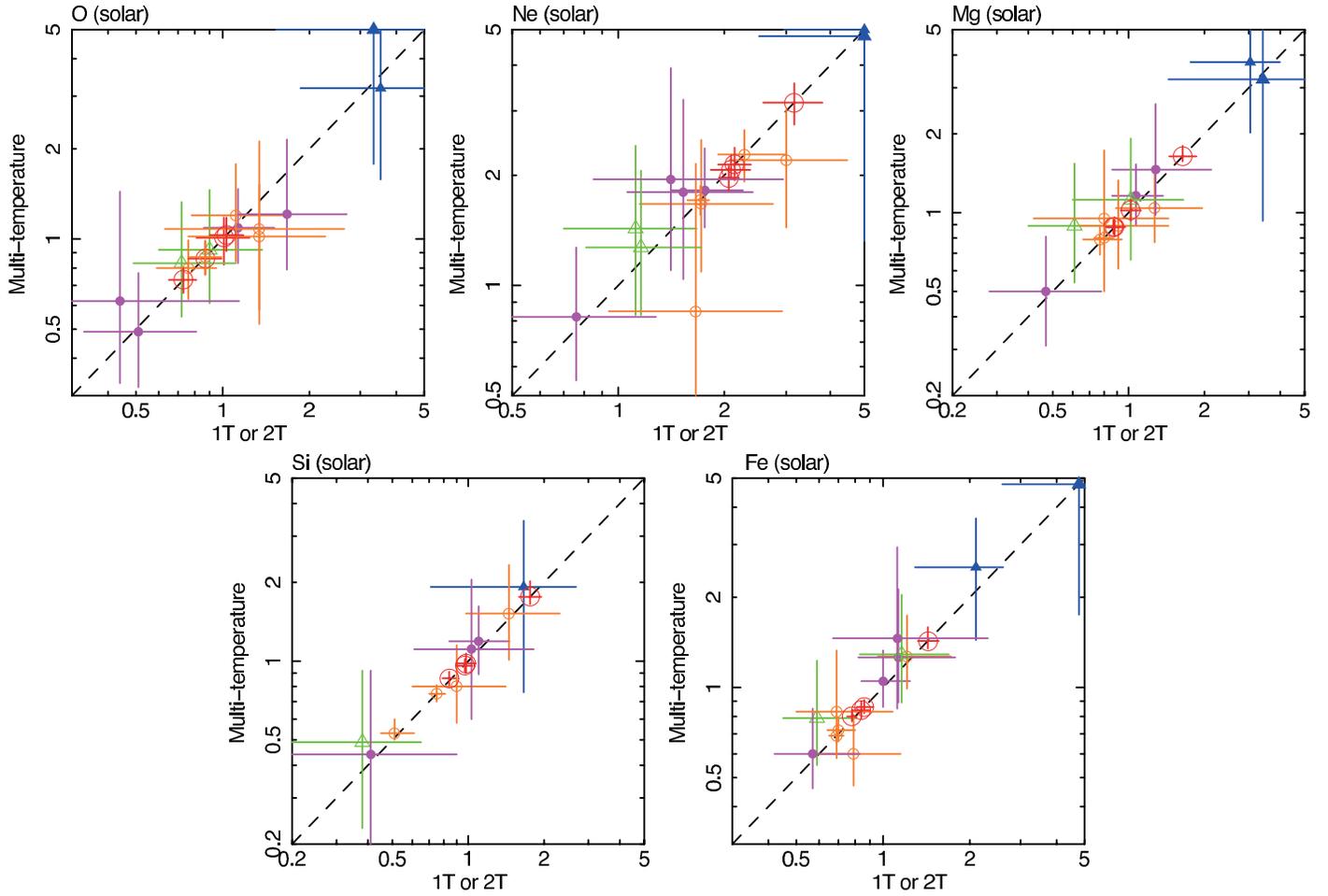}
 \end{center}
 \caption{Comparison of abundances of O, Ne, Mg, Si, and Fe with the multi-T
 model and the 1T or 2T models. 
Statement of meanings of the symbols is shown in Figure \ref {img:kt_O_Mg_Fe_multi}. 
}
\label{img:lodd_apec2_multi_abund}
\end{figure*}

\subsection{Comparison of abundances derived with the 1T or 2T and multi-T  models }

\begin{figure*}[tbp]
 \begin{center}
\includegraphics[width=0.8\textwidth,angle=0,clip]{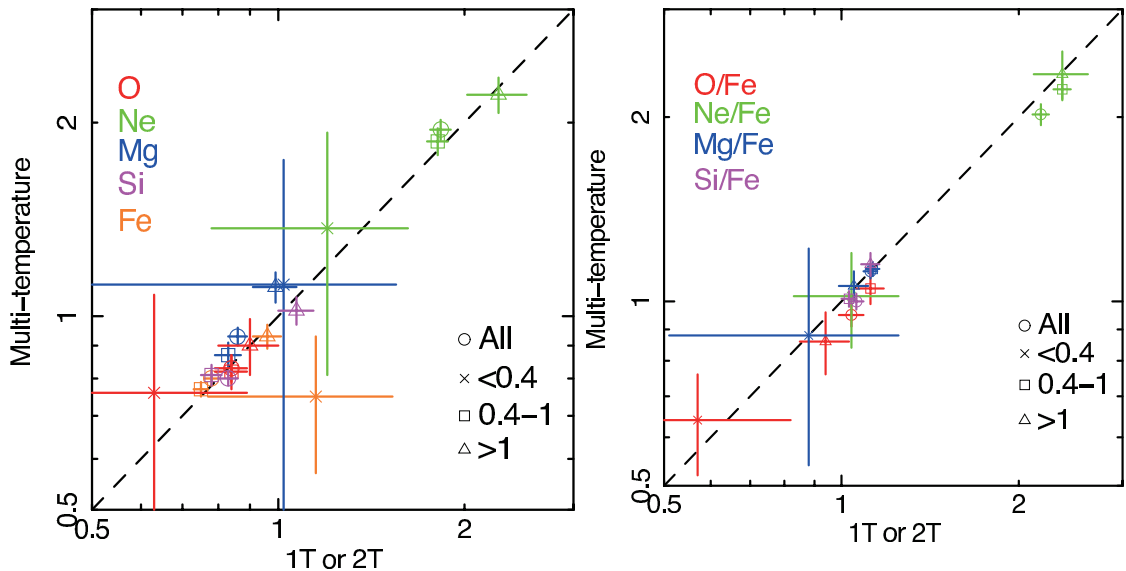}
 \end{center}
 \caption{
 Comparison of the weighted averages of abundances (O, Ne, Mg, Si, and
 Fe; left) and abundance ratios (O/Fe, Ne/Fe, Mg/Fe, and Si/Fe; right)
 with 
the multi-T  and the 1T or 2T models. The
 dashed lines indicate the equal value between two models.
}
\label{img:stat_multi_all}
\end{figure*}

The derived abundances with the multi-T model fits
 are summarized in Table \ref{tbl:metal_multi}.
Figure \ref{img:lodd_apec2_multi_abund} compares the resultant O, Ne, Mg, Si, and Fe
abundances from the 1T or 2T models and multi-T model fits.
For most of the cases, the multi-T model gave almost 
save values except for a few galaxies with  large error bars.
To study the temperature dependence of the differences between the 
abundances from the 1T or 2T and multi-T model fits, we divided galaxies 
into three groups, using the ISM temperature derived from the 1T model
fits of $<0.4$ keV, 0.4--1.0 keV, and $>1.0 $ keV.
Then, we calculated weighted averages of the metal abundances and
metal-to-Fe abundance ratios  derived from 
the 1T or 2T models and the multi-T model fits for galaxies belong to
 each temperature group and the whole sample.
The results are summarized  in 
Figure \ref{img:stat_multi_all} and Table \ref{tbl:mean}.
The weighted averages of
 O, Ne, Mg, Si, and Fe abundances and O/Fe, Ne/Fe, Mg/Fe, and Si/Fe ratios
 derived from the 1T or 2T and the multi-T model fits mostly agree 
well with each other within several \%.
For the galaxies with ISM temperature lower than 0.4 keV, 
the multi-T model tended to yield higher Fe abundances 
than the 1T or 2T models, although the error bars are very large.

\begin{table*}
\begin{center}
\caption{\rm The weighted averages of  abundances of O, Ne, Mg, Si, and
Fe  and O/Fe, Ne/Fe, Mg/Fe, and Si/Fe ratios for all the sample galaxies
and those with ISM temperatures of $<0.4$ keV, 0.4--1.0 keV, $>1.0$ keV.
\label{tbl:mean}}
\begin{tabular}{ccccccccccc}
\tableline\tableline\noalign{\smallskip}
  &  & All &  $<$0.4 keV &  0.4-1 keV &   $>$1 keV     \\       
\noalign{\smallskip}\tableline\noalign{\smallskip}
1T or 2T / v2.0.1 & O&0.84$\pm0.05$&0.63$\pm0.26$&0.85$\pm0.05$&0.90$\pm0.10$\\
  &Ne&1.89$\pm0.07$&1.20$\pm0.42$&1.87$\pm0.08$&2.27$\pm0.25$\\
  &Mg&0.89$\pm0.03$&1.02$\pm0.53$&0.87$\pm0.03$&0.99$\pm0.08$\\
  &Si&0.83$\pm0.03$&\nodata      &0.78$\pm0.03$&1.07$\pm0.07$\\
  &Fe&0.79$\pm0.02$&1.15$\pm0.38$&0.76$\pm0.02$&0.96$\pm0.05$\\

  &O/Fe&1.02$\pm0.05$&0.57$\pm0.25$&1.09$\pm0.06$&0.94$\pm0.09$\\
  &Ne/Fe&2.21$\pm0.07$&1.04$\pm0.21$&2.41$\pm0.08$&2.37$\pm0.25$\\
  &Mg/Fe&1.12$\pm0.03$&0.88$\pm0.37$&1.14$\pm0.03$&1.05$\pm0.06$\\
  &Si/Fe&1.06$\pm0.02$&\nodata      &1.03$\pm0.03$&1.12$\pm0.04$\\

Multi-T / v2.0.1 &O&0.83$\pm0.04$&0.76$\pm0.32$&0.82$\pm0.05$&0.90$\pm0.09$\\
               &Ne&1.95$\pm0.07$&1.37$\pm0.56$&1.87$\pm0.09$&2.21$\pm0.14$\\
               &Mg&0.93$\pm0.03$&1.12$\pm0.63$&0.87$\pm0.04$&1.11$\pm0.06$\\
               &Si&0.80$\pm0.02$&\nodata      &0.81$\pm0.03$&1.02$\pm0.05$\\
               &Fe&0.80$\pm0.02$&0.75$\pm0.18$&0.77$\pm0.02$&0.93$\pm0.04$\\

  &O/Fe&0.95$\pm0.05$&0.64$\pm0.12$&1.05$\pm0.06$&0.86$\pm0.10$\\
  &Ne/Fe&2.02$\pm0.08$&1.02$\pm0.18$&2.22$\pm0.09$&2.35$\pm0.21$\\
  &Mg/Fe&1.12$\pm0.03$&0.88$\pm0.34$&1.13$\pm0.03$&1.06$\pm0.06$\\
  &Si/Fe&1.00$\pm0.03$&\nodata      &1.01$\pm0.03$&1.15$\pm0.05$\\

1T or 2T / v1.3.1 &O & 0.62$\pm0.03$ & 0.40$\pm0.17$ &0.60$\pm0.04$ & 0.87$\pm0.09$ \\
  &Ne& 1.22$\pm0.06$ & 0.78$\pm0.29$  &1.19$\pm0.06$ & 1.84$\pm0.22$ \\
  &Mg& 0.87$\pm0.03$ &0.80$\pm0.42$   &0.82$\pm0.03$ & 1.28$\pm0.11$ \\
  &Si& 0.98$\pm0.03$ & \nodata    &0.93$\pm0.03$ & 1.44$\pm0.09$ \\
  &Fe& 0.94$\pm0.03$ &1.33$\pm0.46$   &0.88$\pm0.03$ & 1.37$\pm0.08$ \\

  &O/Fe&0.60$\pm0.03$& 0.24$\pm0.12$  &0.63$\pm0.03$ & 0.63$\pm0.06$ \\
  &Ne/Fe& 1.09$\pm0.05$ & 0.57$\pm0.12$  &1.20$\pm0.06$ & 1.37$\pm0.17$ \\
  &Mg/Fe& 0.91$\pm0.02$ &0.64$\pm0.27$   &0.91$\pm0.02$ & 0.93$\pm0.04$ \\
  &Si/Fe& 1.00$\pm0.02$ &\nodata     &0.98$\pm0.03$ & 1.06$\pm0.04$ \\
\noalign{\smallskip}\tableline\noalign{\smallskip}
\end{tabular}
\end{center}
\end{table*}


\subsection{Abundances and their ratios with the multi-T model fits}
\label{sec:results}

\begin{figure*}[tbp]
 \begin{center}
\includegraphics[width=\textwidth,angle=0,clip]{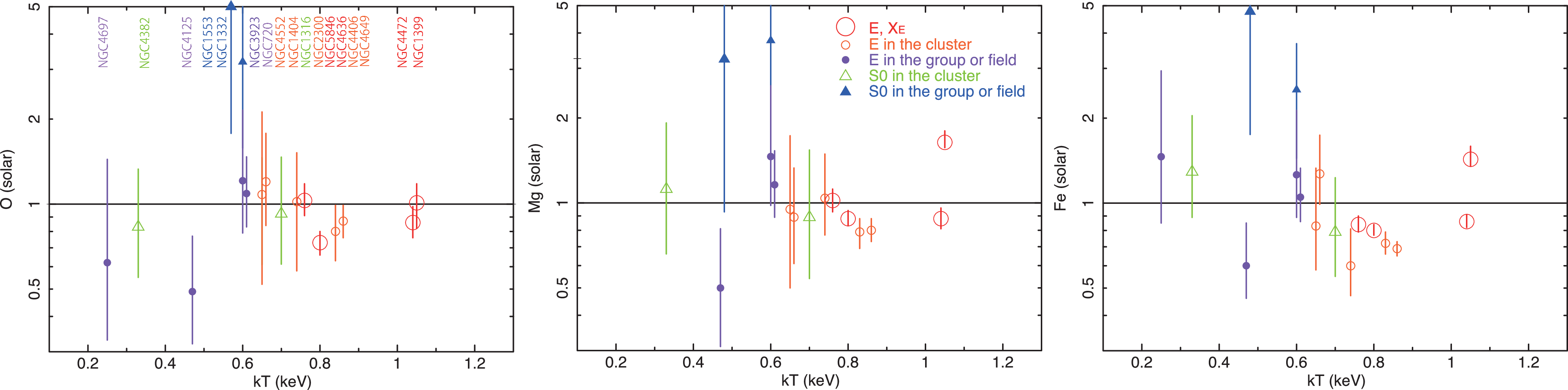}
 \end{center}
 \caption{Abundances of O, Mg and Fe derived from the multi-T model fits. 
Closed and opened triangles (circles) show the abundance patterns of
 S0 (elliptical) galaxies in the field or group 
and cluster, respectively. 
The larger open circles correspond to X-ray
 luminous objects, or $X_{E}$  galaxies \citep{nagino_09}. 
Solid lines represent the solar abundance \citep{lodders_03}.
 }
\label{img:kt_O_Mg_Fe_multi}
\end{figure*}

In Figure \ref{img:kt_O_Mg_Fe_multi}, 
we plotted the  abundances of O, Mg, and Fe derived from the
multi-T model fits against the 1T temperatures of ISM.
The values of O, Mg, and Fe abundances are about 1 solar, 
with no significant dependence  on the ISM temperature.
Elliptical and S0 galaxies have similar values of O, Mg, and Fe abundances.
These abundances  also show no significant dependence on the environment.
The  weighted  averages of the O, Mg, Si, and Fe abundances of all the
 sample galaxies derived from the multi-T model fits are 0.83 $\pm$ 0.04, 
0.93$\pm$0.03, 0.80 $\pm$ 0.02, and 0.80$\pm$0.02 solar, respectively, 
in solar units (Table \ref{tbl:mean}).

\begin{figure*}[tbp]
 \begin{center}
\includegraphics[width=\textwidth,angle=0,clip]{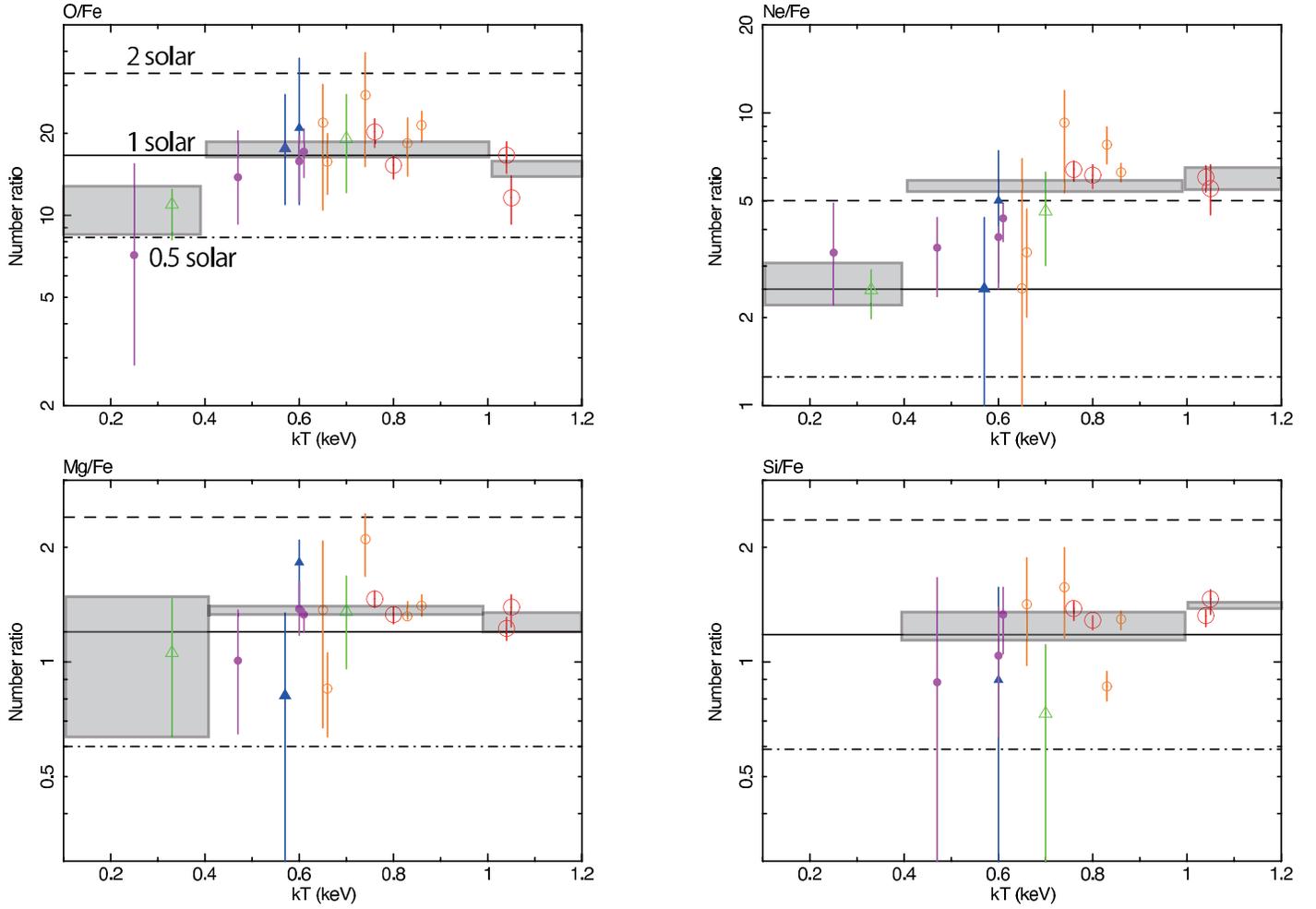}
 \end{center}
 \caption{Abundance ratios of O, Ne, Mg, and Si to Fe derived
from the multi-T model fits.
Meanings of the symbols are the same as those in
 Figure \ref {img:kt_O_Mg_Fe_multi}. 
Solid, dashed, and dot-dashed lines represent the solar abundance \citep{lodders_03}, 
twice, and half of solar ratios, respectively. 
Gray boxes show  the weighted averages for
the three temperature groups, summarized in  Table \ref{tbl:mean}. 
}
\label{img:kt_abundratio_multi}
\end{figure*}

 Figure \ref{img:kt_abundratio_multi} shows the O/Fe, Ne/Fe, Mg/Fe, and Si/Fe
 ratios plotted against the ISM temperature.
These ratios are plotted 
in terms of  number ratios, to avoid differences in different solar abundance tables.
The derived O/Fe, Mg/Fe, and Si/Fe ratios are mostly consistent with the
 solar ratios, except for the O/Fe ratio in the ISM of an S0 galaxy,
 NGC~4382,  with the ISM temperature of 0.33 keV, 
as reported by \citet{nagino_10}.
The weighted averages of the abundance ratios for each temperature group 
are also plotted in Figure \ref{img:stat_multi_all} (right panel).
Those of Mg/Fe and Si/Fe of all temperature groups are close to the solar ratio.
Because of the low O/Fe ratio of NGC~4382, 
the weighted average of the O/Fe ratios of galaxies with ISM temperature below 0.4 keV
is 0.64 $\pm$ 0.12 in unit of the solar ratio, which 
is smaller  than those of the higher temperature groups.
Again, there is no significant dependence on the morphology and environment 
of galaxies.

The derived Ne/Fe ratios are about 2  in unit of the solar ratio,
and NGC~4382 has a significantly smaller Ne/Fe ratio than the other
galaxies by a factor of two (Figure \ref{img:kt_abundratio_multi}).
Since K-shell lines of Ne are  hidden in the Fe-L region, 
the systematic uncertainties in the derived Ne abundances may be significant.
We  discuss the dependence on the Ne abundances on the version of atomic data
at Section \ref{sec:sys}.

\subsection{Effect of the  He abundance in the ISM}

The hot ISM in early-type galaxies are thought to be an accumulation of
stellar mass loss, mainly from asymptotic giant branch (AGB) stars.
As a result, the mass-loss products may have significantly higher He abundances.
From optical observations of AGB stars, the measured He abundances are up to around 2.5 solar \citep{mello_12}. 
Since a  higher He abundance increases the continuum level, 
the assumption of the solar He abundance can yields an underestimation of the Fe abundance.
To study the effect of the assumption of the  He abundance in the ISM,
we fitted the spectra in the same way in Section \ref{ssec:1t2t}, changing the assumed He abundance from 1.2 to 3 solar.
The derived metal abundances have a weak dependence on the He abundance,
and the three solar He abundance gives a 10--30\% higher metal  abundances.
Therefore, systematic uncertainties in the metal abundances caused by
the uncertainty in the He abundance are relatively small.

\section{SYSTEMATIC UNCERTAINTIES BY THE VERSION OF ATOMIC DATA}
\label{sec:sys}

\subsection{Comparison of metal abundances derived from 1T and 2T model}
\label{ssec:abund_1t2t}

\begin{figure*}[tbp]
 \begin{center}
\includegraphics[width=0.5\textwidth,angle=0,clip]{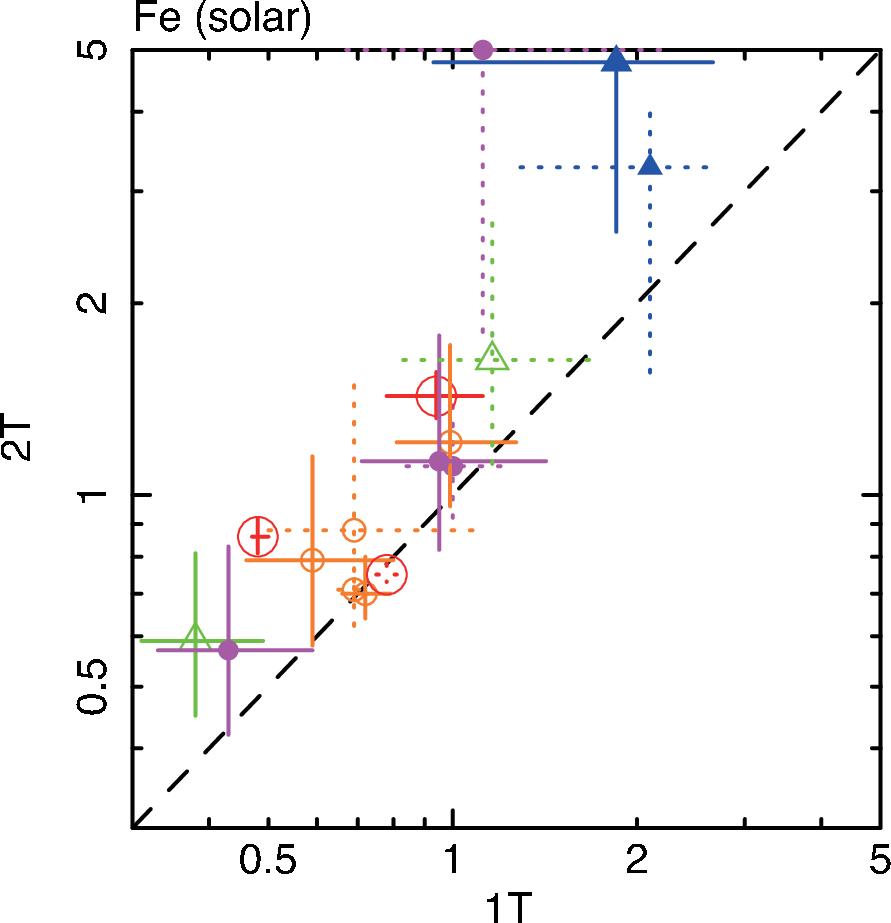}
 \end{center}
 \caption{
Comparison of abundances of Fe with the 1T model and 2T model. 
Statement of meaning of the symbols is shown in Figure \ref{img:kt_O_Mg_Fe_multi}.
The solid and dotted error bars show data, whose best fit model are 2T and 1T model, respectively.
The dashed lines indicates the equal value between two models. 
}
\label{img:lodd_apec2_1tvs2t_Fe_dash}
\end{figure*}

Using {\it ASCA} data, \citet{buote_98} reported that spectral fits with
a multi-temperature plasma model for the ISM in some luminous
early-type galaxies gave larger metal abundances than those with a single temperature model.
We plotted the derived Fe abundances from the 1T and 2T models using AtomDB v2.0.1 
in Figure \ref{img:lodd_apec2_1tvs2t_Fe_dash}. 
The Fe abundances of some galaxies derived from the 2T model fits are higher than those derived
from the 1T model by several tens of \%,
while there are X-ray luminous  galaxies whose Fe abundances derived from the 1T and 2T model fits
are nearly the same.
The differences in the Fe abundances derived from the 1T and 2T model fits
tend to be smaller than those reported by \citet{buote_98}, where a plasma code
of MEKAL \citep{liedahl_95} was used.
The differences in the adopted plasma codes, especially for the Fe-L lines,
also affects the differences in  the derived metal abundances as reported in
previous studies \citep[e.g.,][]{loewenstein_94, matsushita_00, matsushita_07, nagino_10}. 
The hotter temperature component by \citet{buote_98} have temperatures
of a few keV, which may be come from accumulated emission of point sources.

\begin{figure*}[tbp]
 \begin{center}
\includegraphics[width=\textwidth,angle=0,clip]{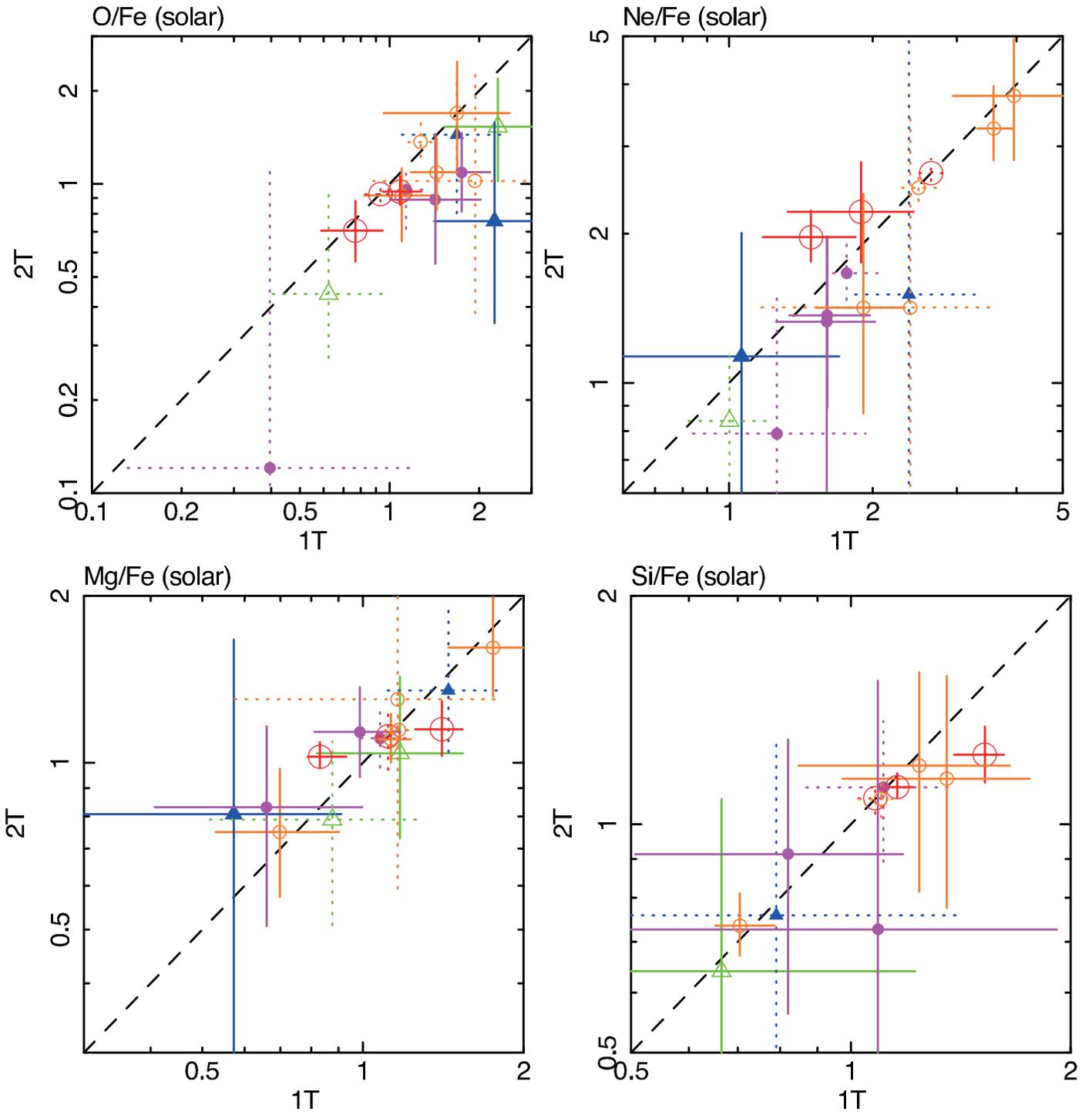}
 \end{center}
 \caption{
Comparison of abundance ratios of O/Fe, Ne/Fe, Mg/Fe, and Si/Fe with the 1T model and 2T model. 
Statement of meaning of the symbols is shown in Figure \ref{img:kt_O_Mg_Fe_multi}.
The solid and dotted error bars show data, whose best fit model are 2T and 1T model, respectively.
The dashed lines indicates the equal value between two models. 
}
\label{img:lodd_apec2_1tvs2t_abundratio_dash}
\end{figure*}

We also plotted the derived O/Fe, Ne/Fe, Mg/Fe, and Si/Fe ratios 
in Figure \ref{img:lodd_apec2_1tvs2t_abundratio_dash}. 
The derived ratios of O/Fe, Ne/Fe, Mg/Fe, and Si/Fe are consistent 
between the 1T and 2T models within error bars, 
although some galaxies have smaller O/Fe ratios from the 2T models compared to those from the 1T models.

\subsection{Comparison of metal abundances derived from AtomDB v1.3.1 and v2.0.1}
\label{ssec:apec}

\begin{figure*}[tbp]
 \begin{center}
\includegraphics[width=\textwidth,angle=0,clip]{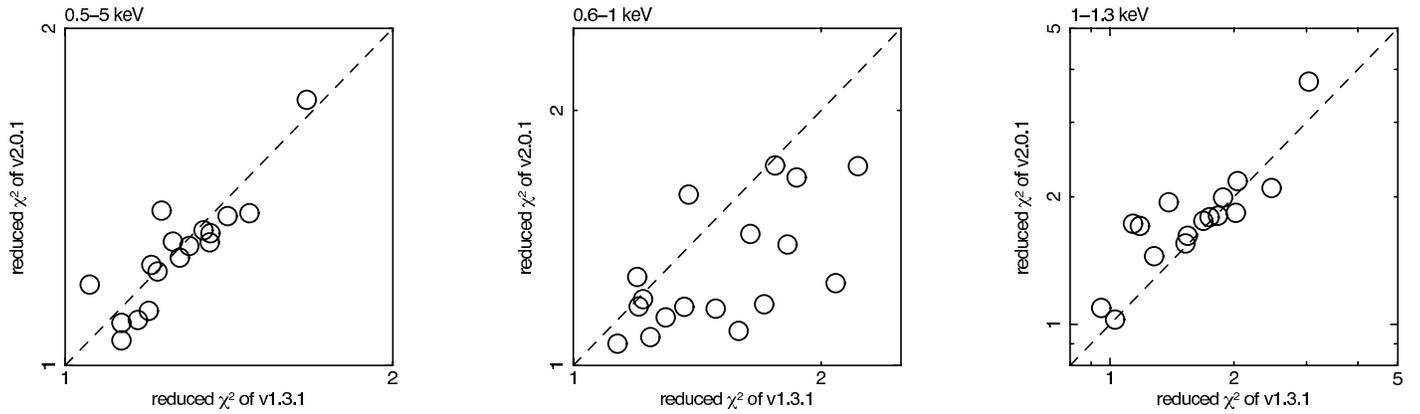}
 \end{center}
 \caption{Comparison of the reduced ${\chi^2}$ derived with AtomDB v1.3.1 and v2.0.1 
in the energy range of 0.5--5 keV, 0.6--1 keV, and 1--1.3 keV in order from left to right panel. 
 }
\label{img:chi_apec12}
\end{figure*}

\begin{figure*}[tbp]
 \begin{center}
\includegraphics[width=0.5\textwidth,angle=0,clip]{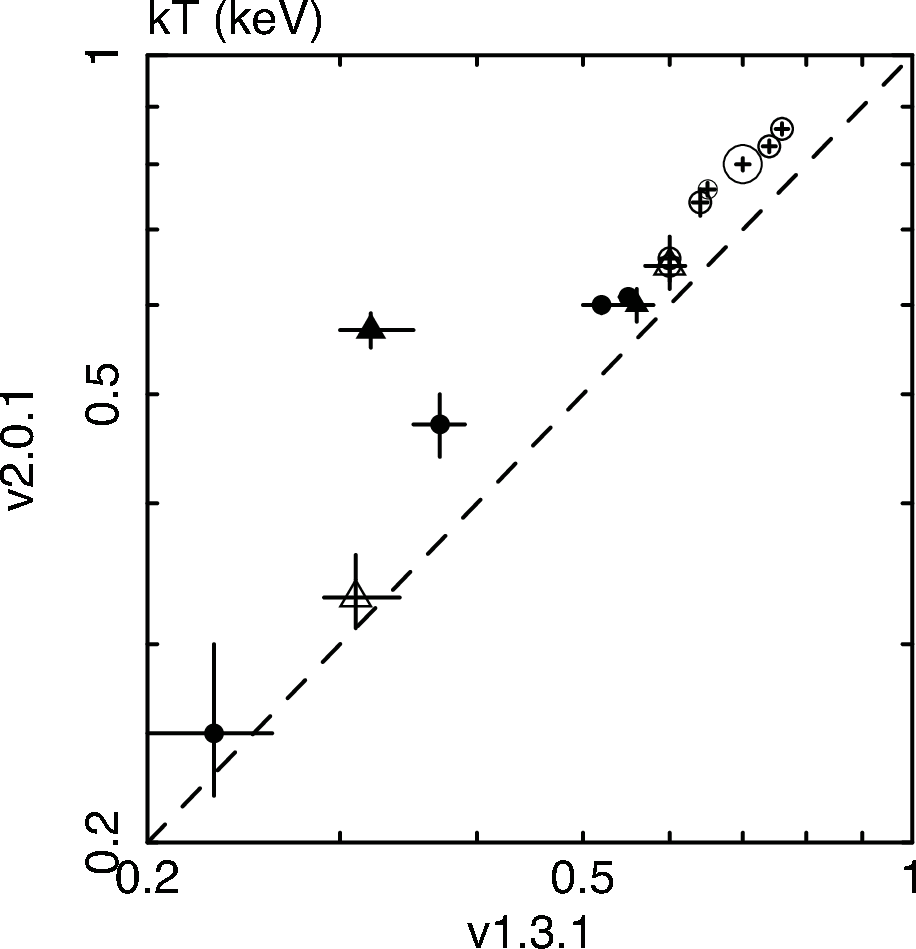}
 \end{center}
 \caption{Comparison of the ISM temperatures derived with AtomDB v1.3.1 and v2.0.1. 
Meanings of the symbols are the same as those in Figure \ref {img:kt_O_Mg_Fe_multi}. 
The dashed line indicates the equal value between two models.
}
\label{img:kt_apec1_apec2}
\end{figure*}

The latest released AtomDB v2.0.1 include several measure update. 
The one of those is improvement in Fe L-shell strength predictions owing to 
incorporating new ionization balance data of Fe.
We performed the spectral fit in the same way with 
the 1T and 2T APEC model for the ISM as in Section \ref{ssec:1t2t}, 
but using AtomDB v1.3.1. 
The residuals of these fits have been listed as bottom panels in Figure \ref{img:spec} (and Appendix \ref{a:analysis}). 
As shown in Figure \ref{img:chi_apec12},
the two versions of AtomDB gave similar reduced ${\chi^2}$  with a some scatter.
The derived temperatures with v2.0.1 are plotted against those with v1.3.1 in Figure \ref{img:kt_apec1_apec2}. 
The temperatures derived with the v2.0.1 code are increased systematically by
$\sim$ 10 \%, except for NGC 1553.
The derived temperatures reflect a change in the peak energy of the Fe-L line blend caused 
by the updates of the Fe-L atomic data.
Since the  emissivity of  Ly$\alpha$ lines has a  temperature dependence, 
the 10\% difference in the derived ISM temperature between the two versions yields
10--20\% under or overestimates of the O, Ne, Mg and Si abundances.

\begin{figure*}[tbp]
 \begin{center}
\includegraphics[width=\textwidth,angle=0,clip]{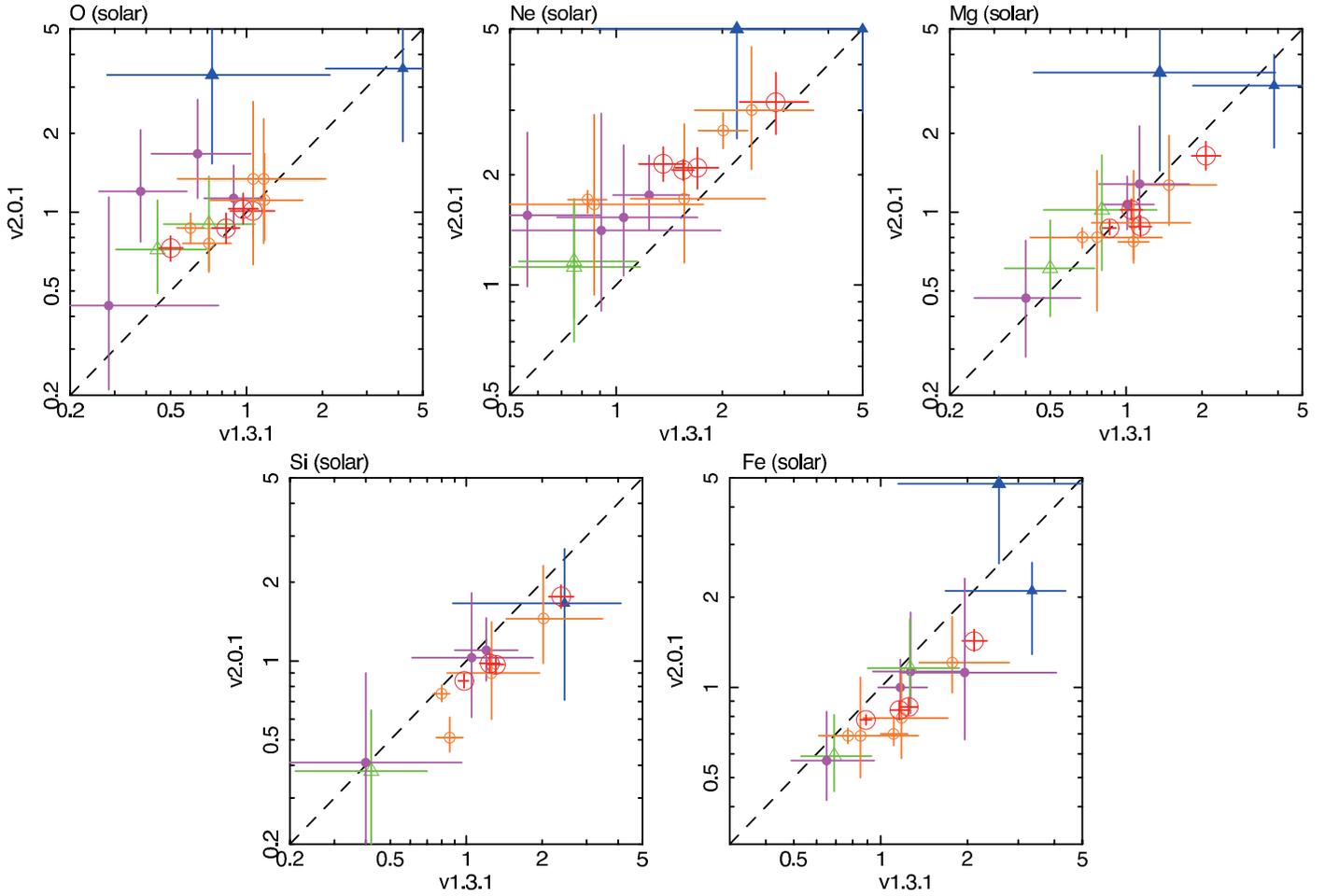}
 \end{center}
 \caption{Comparison of abundances of  O, Ne, Mg, Si, and Fe with v1.3.1 and v2.0.1  models. 
Meanings of the symbols are the same as those in Figure \ref{img:kt_O_Mg_Fe_multi}. 
 }
\label{img:lodd_apec1_apec2_abund}
\end{figure*}

\begin{figure*}[tbp]
 \begin{center}
\includegraphics[width=0.8\textwidth,angle=0,clip]{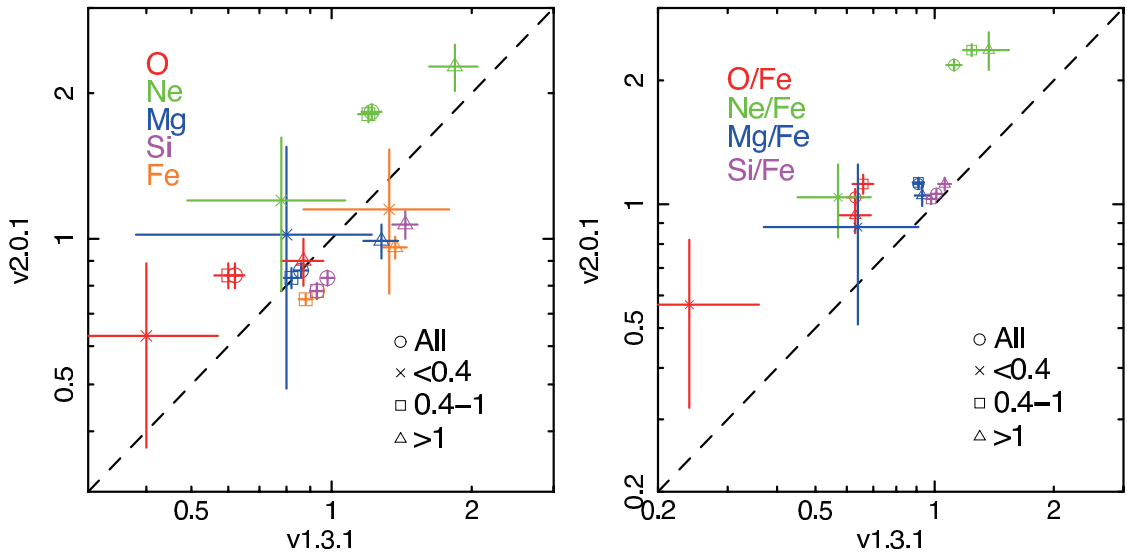}
 \end{center}
 \caption{
Comparison of the weighted averages of abundances (O, Ne, Mg, Si, and Fe; left) and abundance ratios (O/Fe, Ne/Fe, Mg/Fe, and Si/Fe; right) with AtomDB v1.3.1 and v2.0.1. The
 dashed line indicates the equal value between two models.
}
\label{img:stat_apec12_all}
\end{figure*}

Figure \ref{img:lodd_apec1_apec2_abund} 
compare the O, Ne, Mg, Si and Fe abundances 
using the two versions of the AtomDB.
We also calculated the weighted averages of the O, Ne, Mg, Si, and Fe abundances, and
O/Fe, Ne/Fe, Mg/Fe, and Si/Fe ratios for the sample galaxies, and for 
the three temperature groups, and summarized in Table \ref{tbl:mean} and Figure \ref{img:stat_apec12_all}.
The  new version of the AtomDB yielded almost the same values of
 Mg abundances with the old version, while it gave higher O abundances by 30--40\% and
higher Ne abundances by 50\%  and lower Si and Fe abundances by 10--30\%.
As a result,
the O/Fe, Ne/Fe, and Mg/Fe ratios from the version 2.0.1 AtomDB 
are systematically higher than those from the old version by 
a factor of $\sim$ 1.6, 2.0, and 1.2, respectively,
while the Si/Fe ratios did not change.
The difference in the O/Fe ratio tends to be higher for galaxies with lower
ISM temperatures, while no significant dependence on the ISM temperature is 
seen in the differences in the Ne/Fe,  Mg/Fe and Si/Fe ratios.

In \cite{loewenstein_12}, they have also investigated the differences of 
derived temperatures and metal abundances between AtomDB v1.3.1 and v2.0.1, 
although they fitted the simulated spectra, assumed single-temperature 
thermal plasma according to AtomDB v2.0.1. 
Our result, which is the relation of derived temperatures between v1.3.1 and v2.0.1, 
is good agreement with those of \cite{loewenstein_12}. 
Furthermore, they compared the metal abundances, especially Fe, and concluded the derived Fe were overestimated 
using two-temperature models with the v1.3.1, 
which well reproduced the simulated spectra compared to fitting with one-temperature models. 
Our Fe abundances using v1.3.1, whose models are best fit model, are also systematically higher than those of v2.0.1.

\subsection{Residual structures at $\sim$ 0.8 keV}
\label{ssec:0.8kev}

One of the major update in AtomDB v2.0.1 is 
 the strength of emission in energy of 0.7--1.0 keV differ from those of 
v1.3.1, as shown in Figure 3 in \cite{foster_12}. 
The ratio of strong lines of  Fe XVII at 15.0 \AA ~and 17.1 \AA~ 
changed by several tens of \%  from AtomDB v1.3.1 to v2.0.1.
This ratio expected from the latest version of other plasma code, SPEX,
is located between the two versions of AtomDB at a given plasma temperature
\citep{deplaa_12}.
This line ratio is often used to study the effect of  resonant line scattering
to constrain turbulence in the ISM \citep[]{xu_02, hayashi_09,
 werner_09, deplaa_12}.
However, with AtomDB v1.3.1, there are residual structures around 0.8 keV
in the spectra of NGC 1404 and NGC 720 with  ISM temperatures of $\sim$ 0.6 keV 
observed with {\it Suzaku}
\citep{matsushita_07, tawara_08}.
Considering a poor spatial resolution of {\it Suzaku} and relatively compact ISM emission 
of these two galaxies, these residuals reflects the systematic uncertainties in 
the Fe-L atomic data, rather than the resonant line scattering.

As shown in Figure \ref{img:spec} and \ref{aimg:spec1},
using the v1.3.1 AtomDB,
there are strong residual structure around 0.7--0.8 keV in 
the spectra of galaxies with ISM temperature of $\sim$ 0.6--0.8 keV.
These structures mostly disappeared with  v2.0.1. 
In Figure \ref{img:chi_apec12}, the contribution of $\chi^2$ divided by the
number of bins  in the  range of 0.6--1.0 keV using the  v2.0.1 AtomDB
are plotted against those of v1.3.1. 
The $\chi^2$ in this energy range drastically improved with the v2.0.1 AtomDB.
Figure \ref{img:chi_kt_apec12} shows the temperature dependence of ${\chi^2}$/bins
in an energy range of 0.6--0.8 keV.
The improvement of the $\chi^2$ are seen in galaxies with ISM temperatures
of 0.3--0.8 keV.
Since the energy of  Ly$\alpha$ line of  \ion{O}{8} is close the  residual
structure at 0.7--0.8 keV, 
the measurements of O abundance can be affected by the revision of 
the atomic data at this energy range.

\subsection{Residual structures at 1.2 keV}
\label{ssec:1.2kev}

\begin{figure*}[ttbp]
 \begin{center}
\includegraphics[width=\textwidth,angle=0,clip]{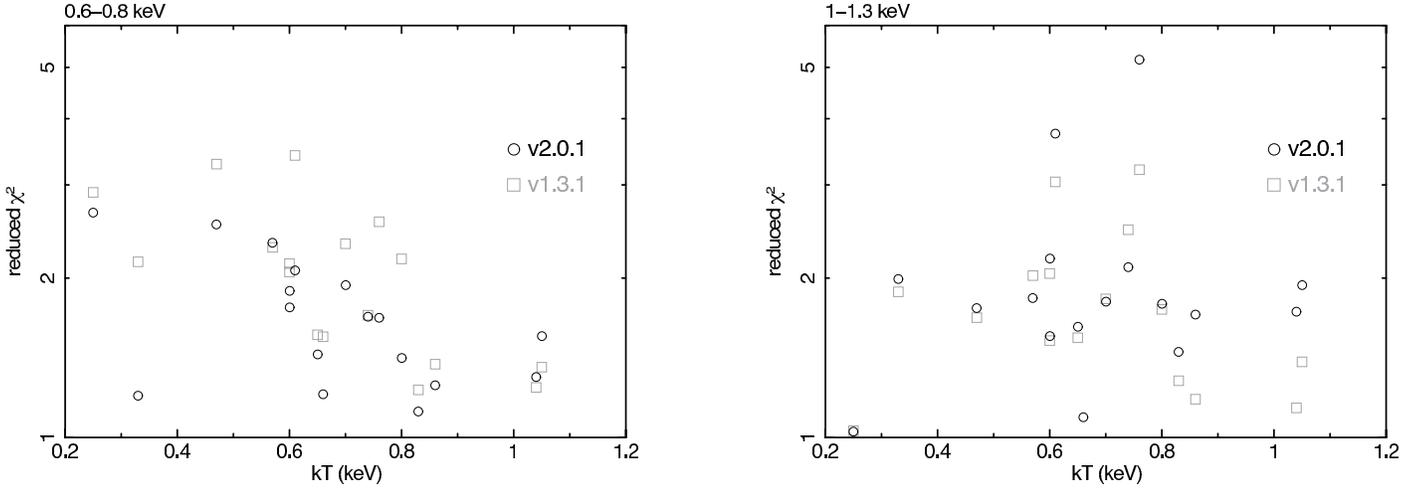}
 \end{center}
 \caption{
The contribution of $\chi^2$ divided by the
number of bins  in the  range of 0.6--0.8 keV (left)
and 1--1.3 keV (right) 
 with AtomDB v1.3.1 (open squares) and v2.0.1 (open circles)
 in the energy range 0.6--0.8 (left) and 
 1--1.3 keV (right), plotted against the temperature of ISM with 1T model. 
}
\label{img:chi_kt_apec12}
\end{figure*}

With the v2.0.1 AtomDB, the contribution of $\chi^2$ divided by the
number of bins  in the energy range of 1--1.3 keV 
of several galaxies increased significantly from the v1.3.1,
although $\chi^2$/d.o.f for most of galaxies are consistent between the two version
(Figure \ref{img:chi_apec12}).
As shown in Figure \ref{img:chi_kt_apec12}, 
some galaxies with the ISM temperatures higher than $\sim$0.6 keV show
higher $\chi^2$ with v2.0.1, while the two versions gave similar 
$\chi^2$ values for the other galaxies with similar ISM temperatures.
These high $\chi^2$ are caused by residual structures around 1.2 keV using
the v2.0.1 AtomDB shown in the
spectra of several galaxies such as NGC 1399 and NGC 4472.
These 1.2 keV residuals  have been observed in various kinds of objects 
\citep[e.g.,][]{yamaguchi_10, loewenstein_12}. 
\citet{brickhouse_00} also interpreted this structure
 as a problem in the Fe-L atomic data. 
These residual structures should be attributed to suppression line
around 1.2 keV in v2.0.1 than those of v1.3.1  from Figure 3 in
\cite{foster_12}.

\begin{figure*}[tbp]
 \begin{center}
\includegraphics[width=\textwidth,angle=0,clip]{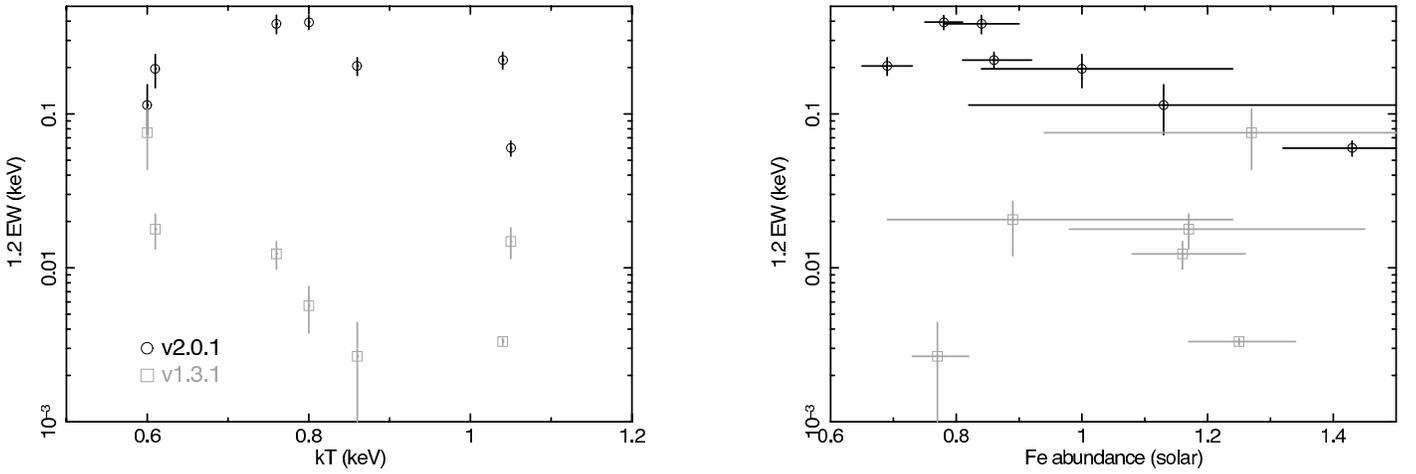}
 \end{center}
 \caption{Equivalent widths of the Gaussian component at 1.23 keV
 are plotted against the temperatures of ISM (left panel) and Fe
 abundance (right panel). The black open circles and gray open squares
 are for  AtomDB v2.0.1 and v1.3.1, respectively.
}
\label{img:fe_apec12}
\end{figure*}

When we added a Gaussian model with the central energy of $\sim$1.23
keV to the spectral models in Section \ref{ssec:1t2t},
the residual structure at $\sim$ 1.2 keV were disappeared.
We plotted the equivalent widths of Gaussian lines from the
two versions,  v2.0.1 and v1.3.1,
for galaxies with the residual structures  against the ISM temperature 
and Fe abundance in Figure \ref{img:fe_apec12}. 
The equivalent widths from v2.0.1 AtomDB are sometimes
 higher by an order of magnitude than those  from v1.3.1.
There are no clear dependence of the equivalent widths on 
the ISM temperature and Fe abundance.
The 1.2 keV residuals should not affect on the O, Mg, Si, and Fe
abundances very much
because the  derived values of these abundances with the Gaussian
component  did not change.
In contrast,  the Ne/Fe decreased. 
Therefore, the larger  Ne abundances with AtomDB v2.0.1 than
those with the old version can be caused by  the change of the Fe-L
line emission around  1.2 keV.

\begin{figure*}[tbp]
 \begin{center}
\includegraphics[width=\textwidth,angle=0,clip]{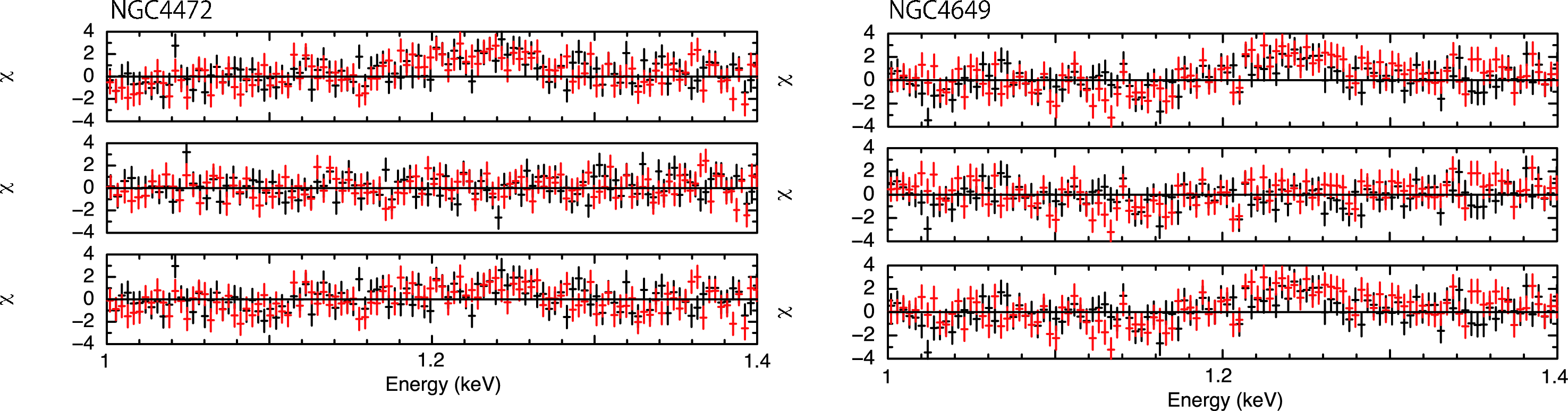}
 \end{center}
 \caption{Residuals of fit the spectra of NGC~4472 (left) and NGC~4649
 (right) using  AtomDB v2.0.1 with the 1T or 2T model
, those with the  Gaussian component at 1.23 keV, those when the
 Ni abundance was allowed to vary in order from upper to bottom panel. 
 }
\label{img:residual_n4649_n4472_1200} 
\end{figure*}

\begin{figure*}[tbp]
 \begin{center}
\includegraphics[width=0.5\textwidth,angle=0,clip]{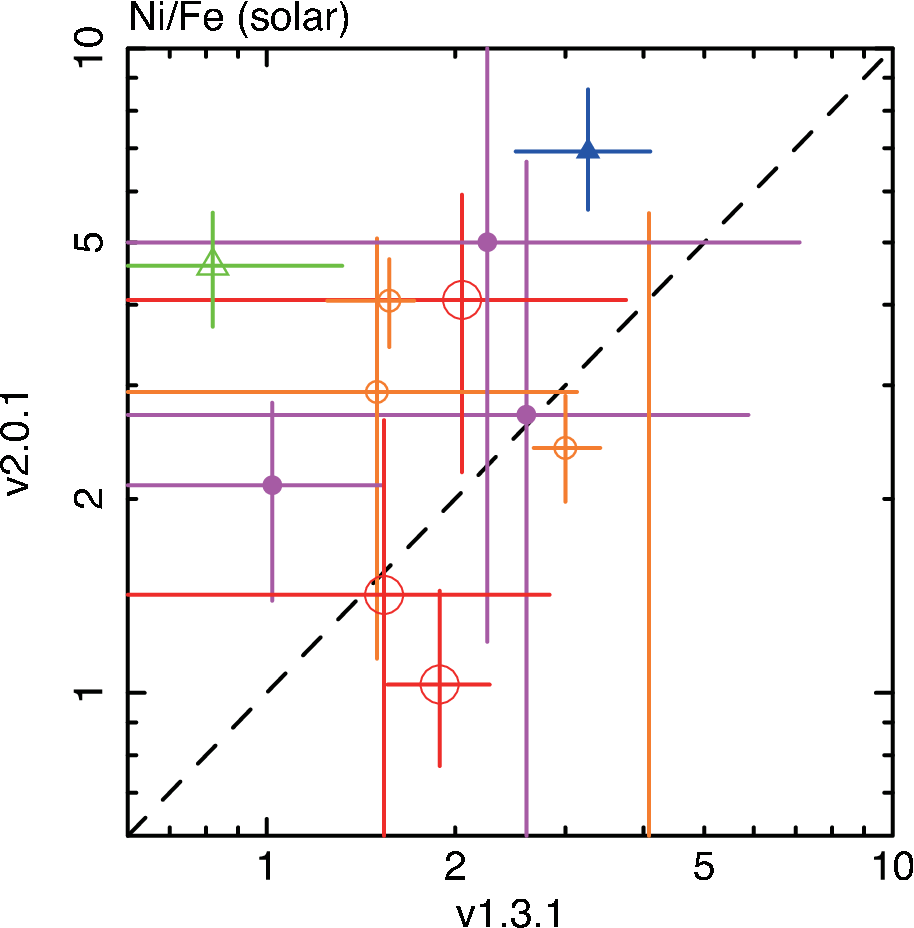}
 \end{center}
 \caption{Comparison of the Ni/Fe ratios derived with AtomDB v1.3.1 and v2.0.1.
The dashed line indicates the equal value between two models.
 }
\label{img:Ni_Fe_apec12_numratio_all}
\end{figure*}

The residuals around 1.2 keV can affect on  the derived Ni abundances,
since the Ni-L lines peak around 1 keV.
We fitted the spectra  with the same model 
as Section \ref{ssec:1t2t} using
the two version of AtomDB, but the Ni abundance was left free. 
However, the residual structures at 1.2 keV still remains as shown in 
Figure \ref{img:residual_n4649_n4472_1200}  (bottom panel).
The  derived Ni/Fe ratios  are plotted in 
 Figure \ref{img:Ni_Fe_apec12_numratio_all}.
The Ni/Fe ratios  become 1 -- 8 in units of the solar ratio,
 with a very large scatter, and the difference in the Ni/Fe ratios
between the two versions of AtomDB  sometimes reaches a factor of 5.
These large Ni/Fe ratios cannot be explained by any nucleosynthesis models
for SN Ia.

In summary, AtomDB v2.0.1 might have a problem in the  Fe-L
emission around 1.2 keV,  
and can affect on the abundance measurements of Ni and Ne abundances.
However, 
the effect on the derived  abundances of Fe, O, Mg, and Si are small.

%
%
%
%

\subsection{Comparison with previous study}
\label{ssec:comparison}

With {\it XMM-Newton} RGS, 
\citet{xu_02}  derived the metal abundances of O, Ne, Fe, and Mg in the
ISM  of an elliptical galaxy, NGC~4636, using an 
old version of APEC plasma code.
With RGS, \citet{werner_09} derived the metal abundances of N, O, Ne,
and Fe in the ISM in 5 elliptical galaxies using SPEX plasma code
\citep{kaastra_96}.
Since RGS has a superior energy resolution, the Ly$\alpha$ of \ion{O}{8} lines 
were clearly detected with RGS.
The derived O/Fe ratios by these papers are about 0.6--0.7 in unit of the solar ratio,
using the solar abundance table by \citet{lodders_03}.
This value agrees well with the weighted average of the O/Fe ratio of
$\sim 0.6$ in unit of the solar ratio of our sample galaxies derived
with the 1T or 2T models for the ISM using AtomDB v1.3.1,
although the new version of AtomDB gave much higher values of the O/Fe
ratio, and absolute values sometimes show discrepancies.

We compare our results of AtomDB v1.3.1  for overlapping objects,
NGC~720, NGC~1399, NGC~3923, NGC~4406, NGC~4472, NGC~4552, NGC~4636, and NGC~4649 
with the previous measurements by \citet{ji_09}, 
which analyzed specific high-quality data of {\it XMM-Newton} EPIC and
RGS and {\it Chandra} ACIS.
The metal abundances by \citet{ji_09} 
are converted using the solar abundance table by \citet{lodders_03}. 
Our Mg/Fe and Si/Fe ratios agree well with those derived by
\citet{ji_09}, while there are discrepancies in the absolute abundances
and O/Fe ratios.


 \citet{loewenstein_10} and \citet{loewenstein_12}, analyzed the metal abundances 
in the ISM of NGC~4472 and NGC~4649 with {\it Suzaku}. 
The results of NGC~4649 are good agreement between their and our study, 
but their metal absolute abundances of NGC~4472 are systematically
larger, although  
our Mg/Fe and Si/Fe ratios agree with those by \citet{loewenstein_10}. 
The major differences for NGC~4472 are version of APEC code,
regions for spectral accumulation, and treatment of background.

In summary, using the same version of plasma code,
our measurements of the O/Fe, Mg/Fe, and Si/Fe ratios  in the ISM
mostly agree with previous measurements,
while the absolute abundances sometimes have discrepancies.
The abundance ratios are derived from the line ratios.
In contrast, 
because the continuum level depends on the value of absolute abundance, 
the difference in the spectra with different absolute abundances is
relatively small.
Therefore, the systematic errors in the absolute abundances can be larger
than the abundance ratios.
Furthermore, the uncertainties 
in the emission from our Galaxy can cause a systematic
error in the derived O abundance.


\section{DISCUSSION}\label{sec:discuss}
\label{sec:discuss}


\subsection{Metal abundance patterns and contributions from SNe}
\label{ssec:metal}

\begin{figure*}[tbp]
 \begin{center}
\includegraphics[width=0.4\textwidth,angle=0,clip]{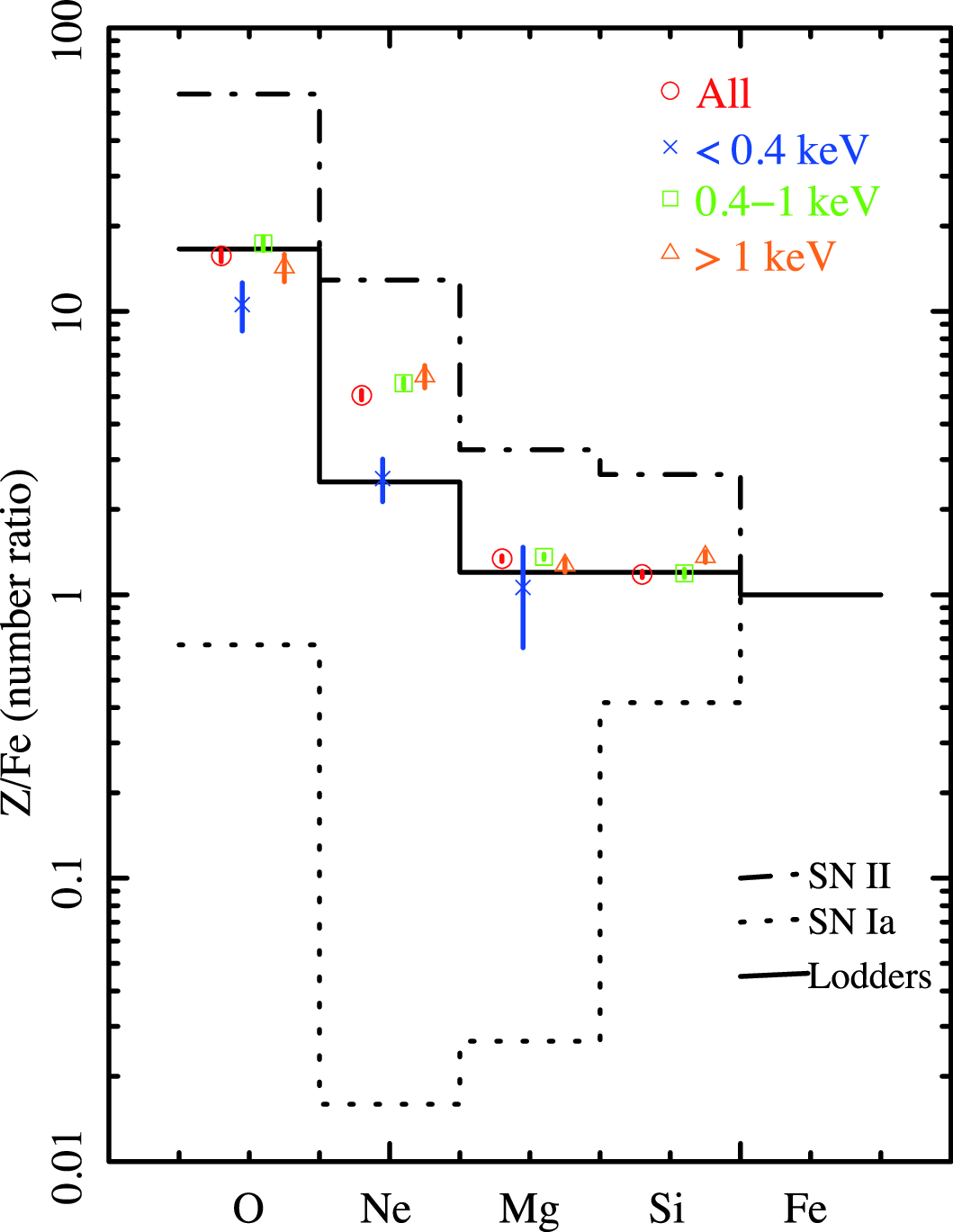}
 \end{center}
 \caption{The weighted averages 
of the  abundance ratios of O, Ne, Mg, and Si to Fe for the multi-T
 model with AtomDB v2.0.1. 
Red, blue, green, and orange data show for all the sample galaxies, 
temperature groups of  $<$0.4 keV, 0.4--1 keV, 
and $>$1 keV, respectively.
 Solid, dot-dashed, and dotted lines represent the number ratios of metals to Fe for solar abundance \citep{lodders_03}, for SNcc products \citep{nomoto_06}, and for SN Ia ones \citep{iwamoto_99}, respectively. 
}
\label{img:num_ratio_all_04_04_1_1} 
\end{figure*}

We successfully measured the abundance patterns of O,  Mg, Si, 
and Fe in the ISM of 17  early-type galaxies,  13 ellipticals and 4 S0's,  with $L_K>L_{K*}$    and derived abundance ratios. 
Figure \ref{img:num_ratio_all_04_04_1_1} shows the weighted averages of
the  O/Fe, Ne/Fe, Mg/Fe, and Si/Fe ratios  of all of the sample galaxies
derived with the multi-T fits using the APEC plasma code with AtomDB v2.0.1.
The abundance ratios except for the Ne/Fe ratio
agree very well with  the solar  ratios by \citet{lodders_03}.
The yields of core-collapse SN (hereafter SNcc) and SN Ia  are also plotted in Figure
\ref{img:num_ratio_all_04_04_1_1}.
Here, the SNcc yields by \citet{nomoto_06} refer to an 
average over the Salpeter initial mass function of 
stellar masses from 10 to 50 $M_{\odot}$, with a progenitor metallicity of $Z=0.02$\@.
The SNe Ia yields  were taken from the W7 mode by \citet{iwamoto_99}.
The abundance ratios in the ISM are located between those of SNcc and
SN Ia. 
The  solar  O/Fe, Mg/Fe, and Si/Fe ratios in the ISM indicate
a  same mixture of the two types of SN  with the solar system.
Considering that 
the O/Fe and Mg/Fe ratios of the  SNcc yields expected from  
theoretical nucleosynthesis models
by \citet{nomoto_06} and observed abundance pattern of 
metal poor stars in our Galaxy \citep[e.g.,][]{edvardsson_93, feltzing_98, bensby_03, bensby_04} 
are about three in units of the solar ratio,
 $\sim$70\% and $\sim$30\% of
 Fe are synthesized by SN Ia and SNcc, respectively,

The observed Ne/Fe ratios, about two in unit of the solar
ratio, and the large Ni/Fe ratios with a significant scatter,
cannot be explained by any mixture of the two types of SNe
which can  reproduce the solar O/Fe, Mg/Fe, and Si/Fe ratios.
The Ne and Ni abundances  may have intrinsic large systematic errors 
because their emission lines are  hidden by prominent Fe-L lines.

The weighted averages of the abundance ratios of the hotter two
temperature groups, 0.4--1 and $>1$ keV, agree very well with those for all the sample
galaxies (Figure \ref{img:num_ratio_all_04_04_1_1}).
On the other hand,  mainly due to the low O/Fe and Ne/Fe abundances in the ISM 
of  a S0 galaxy, NGC~4382, the lowest temperature group has smaller
O/Fe and Ne/Fe ratios by a factor of two.
To explain the lower O/Fe and Ne/Fe ratios in the ISM for the lowest
temperature groups, we need a higher contribution of
SNe Ia to Fe in the ISM.

\subsection{Comparison with optical measurements of stellar metallicity}
\label{ssec:stellar_metal}

\begin{figure*}[tbp]
 \begin{center}
\includegraphics[width=0.8\textwidth,angle=0,clip]{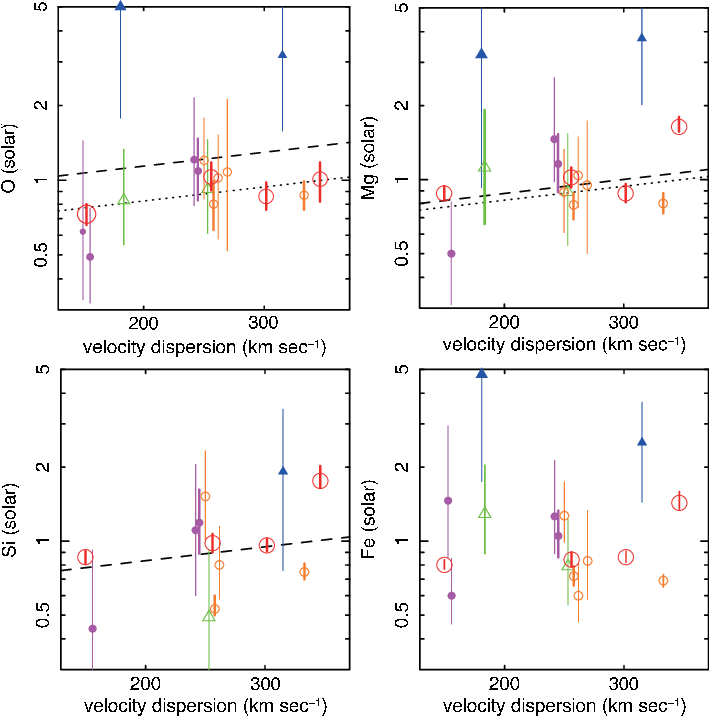}
 \end{center}
 \caption{The abundances of O, Mg, Si, and Fe in the ISM derived from the
 multi-T model fits with AtomDB v2.0.1
are plotted against 
the central stellar velocity dispersion by \citet{prugniel_96}. 
Meanings of the symbols are the same as those in Figure \ref {img:kt_O_Mg_Fe_multi}. The dashed and dotted lines 
show the $Z$--$\sigma$ relation for a $r_{\rm e}$ aperture derived
from  optical spectroscopy  by \citet{kuntschner_10} and that converted considering the difference in the solar abundance tables, respectively. 
 }
 \label{img:sigma_abund_multi_solar_dis}
\end{figure*}

Figure \ref{img:sigma_abund_multi_solar_dis} shows the O, Mg, Si, and Fe abundances 
in the ISM derived from the multi-T model using the version 2.0.1 AtomDB, 
plotted against the central stellar velocity dispersion, $\sigma$.
As found by optical observations for the stellar metallicity \citep[e.g.,][]{kuntschner_10}, 
there is no environmental and morphological dependence on the O, Mg, and Si abundances.
In Figure \ref{img:sigma_abund_multi_solar_dis}, 
we also plotted the  best-fit relation of the stellar 
metallicity, $Z$, within $r_{\rm e}$ against $\sigma$ derived from optical spectroscopy 
by \citet{kuntschner_10}.
Since O and other $\alpha$-elements mostly contribute the metallicity, 
we compared $Z$ with O, Mg, and Si abundances in the ISM \citep[e.g.,][]{tantalo_98}.
Among $\alpha$ elements, Mg abundances in the ISM have smallest systematic error 
in our analysis and are mainly used to derive metal abundance $Z$ in optical observations.
Except for a few galaxies, the derived Mg and Si abundances in the ISM mostly 
agree with the $Z$--$\sigma$ relation by
\citet{kuntschner_10}.
Although Si are synthesized by both SN Ia and SNcc, 
the solar abundance pattern indicates that most of Si
come from SNcc and therefore, it is reasonable to have the same $Z$--$\sigma$ relation with Mg.
Two S0 galaxies in the field or small groups show significantly higher Mg abundances 
than the other galaxies. However, the systematic differences due to different versions 
of APEC models and temperature modelling of the ISM in these galaxies are 
also larger than the other ones.

In \citet{kuntschner_10}, they adopted the solar abundance table by \citet{grevesse_98},
and the other for the abundance pattern of CN-strong stars \citet{canon_98}.  
In optical observations, the metallicity is derived from the combination of the strength 
of Mg and Fe absorption lines.
However,  the solar O abundance derived from solar photospheric lines has been changed 
by several tens of \%,
considering three-dimensional hydro-static model atmospheres and non-local thermodynamic equilibrium
\citep[e.g.,][]{asplund_05}. In this paper, we use the new solar abundance table 
by \citet{lodders_03},  adopting this O abundance.
Therefore, in Figure \ref{img:sigma_abund_multi_solar_dis}, 
we plotted the same $Z$--$\sigma$ relation by \citet{kuntschner_10} 
and that
converted to the table by \citet{lodders_03} considering the difference in the solar abundance tables.
The O abundances in the ISM agree well with the original $Z$--$\sigma$ relation, 
but systematically offset
from the converted one.
However, the slope of the O abundance--$\sigma$ relation agrees very well with 
the optical $Z$--$\sigma$ relation.
When we compared the absolute metal abundances between optical and X-ray observations, 
there are systematic errors coming from the differences of solar abundance tables, 
the ways of the observations, and uncertainties of emission models. The trends of
all elements are agree with the $Z$--$\sigma$ relation.


Optical observations indicate that negative radial  gradients of stellar metallicity
in elliptical galaxies are common.
Assuming that the same metallicity gradient continues beyond
$r_{\rm e}$,  the average stellar metallicity of entire galaxy is similar
to that at $r_{\rm e}$ \citep{kobayashi_99}.
The differences in the stellar metallicity within $r_{\rm e}$ and
at $r_{\rm e}$ of elliptical galaxies derived by the optical observations
 are typically around 0.1 dex, or a factor of 1.2--1.3
\citep{kuntschner_10}.
Our measurements of the ISM abundances indicate that
the  average stellar metallicity of entire galaxies with 
$\sigma > 200 \rm{km s^{-1}}$ are close to the solar metallicity,
and the  metallicity-mass relation is consistent with that 
of  the stellar relation for the $r_{\rm e}$ region.
Although 
we are comparing abundances in two distinct media,
 stars and ISM, which could have very different histories, and 
different systematic errors,  
the agreement of the metallicity between the hot ISM and
stars indicates relatively small systematic errors in the measurements
with these optical and X-ray observations.


\subsection{Comparison with optical measurements of stellar
   $\alpha$/Fe ratios}

\begin{figure*}[tbp]
 \begin{center}
\includegraphics[width=\textwidth,angle=0,clip]{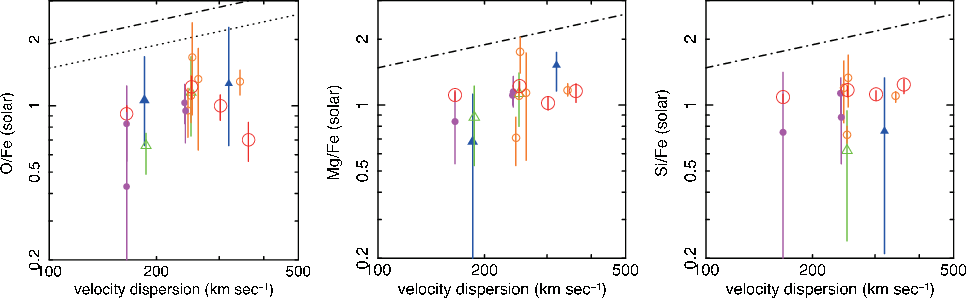}
 \end{center}
 \caption{The abundance ratios of O/Fe, Mg/Fe, and Si/Fe in the ISM
derived from the multi-T model fits with AtomDB v2.0.1 are 
plotted against central stellar  velocity dispersion by \citet{prugniel_96}. 
Meanings of the symbols are the same as those in Figure \ref {img:kt_O_Mg_Fe_multi}. 
The dot-dashed and dotted lines  show the [$\alpha$/Fe]--$\sigma$ relation 
for a $r_{\rm e}$ aperture derived
from  optical spectroscopy  by \citet{kuntschner_10} and that converted considering the difference in the solar abundance tables, respectively.
 }
 \label{img:sigma_abundratio_multi_solar}
\end{figure*}

In Figure \ref{img:sigma_abundratio_multi_solar}, 
the O/Fe, Mg/Fe, and Si/Fe ratios in the ISM are plotted against $\sigma$.
We note that the Mg/Fe ratios have smallest systematic differences
between the two versions of plasma codes and among the different modeling
of the temperature structure for the ISM.
Again, there is no systematic difference in the O/Fe, Mg/Fe, and Si/Fe ratios
between the ellipticals and S0 galaxies, or between those in
the field or small groups and in clusters.
The O/Fe and Mg/Fe ratios of the two S0 galaxies with 
higher O and Mg abundances are consistent with those of the others.
The dot-dashed lines in these plots correspond to the best-fit
relation between [$\alpha/\rm{Fe}$] within $r_{\rm e}$  and $\sigma$ 
derived from optical spectroscopy by \citet{kuntschner_10}.
The O/Fe, Mg/Fe, and Si/Fe ratios in the ISM are systematically smaller
by a factor of two  than the relation for the  stellar metallicity.
Therefore, we need additional SN Ia enrichment in the ISM to account for
both of the X-ray and optical observations of early-type galaxies.

A longer star-formation time scale yields more SNe Ia products
in stars, and therefore, the difference in the $\alpha$/Fe ratio
in stars can constrain the star formation histories.
As shown in Figure \ref{img:sigma_abundratio_multi_solar},
the optical observations indicate that massive galaxies show 
higher O/Fe, Mg/Fe, and Si/Fe ratios and therefore shorter  star formation time
scales.
Considering typical
timescale for SN Ia from star formation \citep{kobayashi_09},
 duration of major  star formation should be shorter  than a few Gyr.
Although we cannot distinguish the SNe Ia yields in the ISM from
present SNe Ia and those trapped in stars 
and the errors in the O/Fe, Mg/Fe, and Si/Fe ratios in the ISM 
of galaxies with small $\sigma$  are relatively large,
our measurements are consistent with the [$\alpha$/Fe]--$\sigma$ relation 
derived from the optical spectroscopy by \citet{kuntschner_10} for
a $r_{\rm e}$ aperture,
assuming an additional enrichment of the Fe abundance of $\sim$ 0.5
solar from the present SN Ia.
As described in Section \ref{ssec:stellar_metal}, although both SN Ia an SNcc produce Si, 
the Si/Fe ratios have similar trend compared to those of O/Fe and Mg/Fe ratios.
Since the Si/Fe ratio is solar abundance, Si in the ISM should come from by SNcc. 


There are no difference in the abundances and their ratios in the ISM
between the cluster galaxies and those in the field or small groups.
The similarity of the abundances in the ISM indicate that
major star formation history are similar.
There are also no dependence in the ISM abundance patterns 
between elliptical and S0 galaxies. 
Then, major star formation histories 
 of these galaxies may be similar with those of ellipticals,
although a possibility of spiral galaxies changed into S0 galaxies
is discussed based on the 
 fractional evolution of S0 and spiral galaxies in clusters
\citep[e.g.,][]{dressler_97, kodama_04, poggianti_09}. 
The two S0 galaxies in clusters in our sample, NGC~1316 and NGC~4382 
have indications of experiences of recent  major mergers
\citep[e.g.,][]{goudfrooij_01, sansom_06}.
NGC~4382  has  smaller O/Fe and Ne/Fe ratios in the ISM by a factor of two
than the other galaxies.
Considering that the reported present SN Ia rate in S0 and elliptical galaxies
are consistent with each other \citep[e.g.,][]{mannucci_08},
Fe abundances from recent SN Ia should be similar between the two types of galaxies.
Consequently, the SN Ia products included in stars of NGC~4382 are larger than 
those of other early-type galaxies. 
If this galaxy is formed with a merging of massive spiral galaxies,
the stars contains significant amount of SN Ia yields as in our Galaxy, 

\subsection{Present Fe enrichment by SN Ia}
\label{ssec:fe_metal}

The Fe abundance synthesized by present SNe Ia in an early-type 
galaxy is calculated by  $M^{\rm Fe}_{\rm SN}\theta_{\rm SN}/\alpha_{*}z^{\rm Fe}_{\rm solar}$ 
(see \cite{matsushita_03} for details). 
Here, $M^{\rm Fe}_{\rm SN}$ is the Fe 
mass synthesized by one SN Ia, $\theta_{\rm SN}$ is the SN Ia rate, 
$\alpha_{*}$ is the stellar mass loss rate, and $z^{\rm Fe}_{\rm solar}$ is the solar Fe mass fraction. 
We used the mass-loss rate from \citet{ciotti_91}, which is approximated by $1.5 \times 10^{-11} L_{\rm B}
t^{-1.3}_{15}M_{\odot}\rm{yr}^{-1}$, where $t_{15}$ is the age in units of 
15 Gyr and $L_{\rm B}$ is the B-band luminosity. $M_{\rm Fe}$ produced by one 
SN Ia explosion is likely to be $\sim 0.6M_{\odot}$ \citep{iwamoto_99}. 
Since there are no significant difference of SN Ia rate between elliptical and S0 galaxies, 
we adopted 0.1--0.5 $\rm{SN~Ia}/100\rm{yr}/$ $10^{10}L_{\rm B}$ as the 
optically observed SN Ia rate 
\citep[e.g.,][]{mannucci_08, blanc_04, hardin_00, cappellaro_97}. 
We used the solar Fe mass fraction 0.0012 from \citet{lodders_03}. 
The estimated stellar age assuming a single-stellar population 
for massive early-type galaxies are typically older than 10 Gyr
\citep[e.g.,][]{thomas_05, kuntschner_10}.
The resultant Fe abundance is 2.8-13.9 solar assuming the stellar age of
13 Gyr.
Assuming a stellar age of 10 Gyr, 
the mass-loss rate increase by 1.4 times, 
and the contribution of SN Ia to Fe abundance decreases by 0.7 times,
and the expected Fe abundance from SN Ia becomes 2.0--9.7 solar.

Since the weighted average of the Fe abundances in the ISM is 0.8 solar, 
the contribution from SN Ia should be about 0.5 solar 
($\sim$70\% of Fe abundance synthesized by SN Ia from Section \ref{ssec:metal}), 
which is a sum of those in stars and present SN Ia.
On the other hands, if an average of the stellar metallicity and  $[\alpha $/Fe] ratio over entire galaxies 
are close to the relations of $Z$--$\sigma$ and [$\alpha$/Fe]--$\sigma$, respectively,
derived by \citet{kuntschner_10}, 
we need a Fe abundances of about $\sim$ 0.5 solar from present SNe Ia 
(Figure \ref{img:sigma_abundratio_multi_solar}).
Therefore, 
if all the  ejecta of SNe Ia have been completely  mixed into the ISM,
the  present SN Ia rate to account for the observed Fe abundance in the  ISM
is significantly smaller than those measured by optical SN Ia observations. 
With estimated Fe abundance from present SNe Ia, 
the calculated SN Ia rate is $\sim$0.02.

If some part of  SN Ia ejecta can escape the ISM
before fully mixed in to the ISM \citep{matsushita_00, tang_10,
loewenstein_12}, the Fe abundance can be lower.
\citet{tang_10} simulated the evolution of SN Ia ejecta in the galaxy-wide hot
gas out-flows. They found that SN Ia ejecta producing little X-ray emission and
driven by its large buoyancy, can quickly get higher outward velocity.
Since this ejecta slowly diluted and cooled, and as a result,
they expect that the emission-weighted Fe abundance of central few kpc
becomes significantly smaller.
However, no significant radial gradients in the Mg/Fe ratios are detected
with {\it Suzaku}, {\it XMM} and {\it Chandra} observations \citep{ji_09, hayashi_09,
loewenstein_12}.
Furthermore, the X-ray luminous objects surrounded by
larger-scale potentials (large open circles in Figure
\ref{img:sigma_abund_multi_solar_dis} and
\ref{img:sigma_abundratio_multi_solar}), 
and NGC~1404, which has a compact ISM emission due to ram-pressure
stripping \citep{machacek_05},
have similar values of O/Fe and Mg/Fe ratios 
as well as the O, Mg, and Fe abundances with the other galaxies.
Since the ISM mass within 4 $r_{\rm e}$ in these X-ray luminous galaxies
are several times larger than those in the X-ray fainter galaxies,
the accumulation time scale  for the ISM should be different.
Therefore, the idea of  the partial mixing of the SN Ia ejecta into the
ISM also has an difficulty to explain the observed abundance pattern
in the ISM.

The ICM in clusters of galaxies contains a large amount of metals.
The observed abundance pattern of the ICM indicates that most of Fe
in the ICM were synthesized by SN Ia \citep[e.g.][]{Sato_07, Sato_09, Matsushita_13}.
{\it Suzaku} enabled us to measure the Fe mass in the ICM out to the virial radius
\citep{Sato_12, Matsushita_13}.
The observed ratios of Fe mass in the ICM to the total light from galaxies, 
iron-mass-to-light ratio (IMLR), out the virial radius of Hydra A and the Perseus 
clusters reach $\sim 10^{-2} M_\odot/L_{K,\odot}$.
However, accumulating the observed SN Ia rate by optical observations
over the Hubble time, 13.7 Gyr, 
the expected IMLR from the SN Ia 
becomes (2--5)$\times 10^{-4} M_\odot/L_{K,\odot}$.
Adopting the SN Ia rate  indicated from X-ray measurements of the Fe abundance
in the ISM assuming that SN Ia ejecta well mixed  into the ISM,
the expected IMLR from the SN Ia becomes even smaller,  $\sim 10^{-4} M_\odot/L_{K,\odot}$.
These results indicate that
the lifetimes of most of SN Ia are much shorter than
the Hubble time, and the SN Ia rate in cluster galaxies was much higher in the past.

\section{CONCLUSION}
\label{sec:con}

We performed X-ray spectral analysis of 17 early-type galaxies,
13 ellipticals and 4 S0s, with {\it Suzaku}. 
The spectra extracted from 4 $r_{\rm e}$ 
have been produced with 1T or 2T thermal 
plasma model and the multi-T model for the ISM.
We successfully measured the metal abundance patterns O,  Mg, Si, and Fe
of the ISM  and derived abundance ratios. 
The  weighted  averages of the O,  Mg, Si, and Fe abundances of all the
 sample galaxies derived from the multi-temperature model fits are 0.83 $\pm$ 0.04, 
0.93$\pm$0.03, 0.80 $\pm$ 0.02, and 0.80$\pm$0.02 solar, respectively, 
in solar units according to the solar abundance 
table by \citet{lodders_03}. 
The abundance ratios of O/Fe, Mg/Fe, and Si/Fe are close to the solar
ratio.
The derived values of  Ne and Ni abundances may have larger systematic 
uncertainties because their emission lines are hidden by the Fe-L lines.
The O and Mg abundances in the ISM within 4 $r_{\rm e}$ agree well with
the stellar metallicity derived by the optical observations for $r_{\rm
e}$ apertures.
This agreement  indicates relatively small systematic 
errors in the measurements with these optical and X-ray observations.
The solar O/Fe and Mg/Fe ratios in the ISM indicate additional
contribution from present SN Ia.
There is no systematic differences between galaxies in clusters
and field or small groups or between elliptical and S0 galaxies.
Therefore, major star formation history should be similar among these
objects.
The Fe abundance in the ISM is significantly smaller than
the expected value derived from optical observations,
indicates a low present SN Ia rate.

\acknowledgments
We thank the referee for providing valuable comments. 
We would like to thank Tadayuki Kodama for valuable comments. 
We gratefully acknowledge all members of the {\it Suzaku} hardware and software teams and the Science Working Group. 
SK is supported by JSPS Research Fellowship for Young Scientists



\appendix

\section{TREATMENT AND RESULTS OF CONTAMINANT ON XIS}
\label{a:contami}

The quantum efficiency of XIS in the low energy range 
has been decreasing owing to OBF (see detail in The Suzaku Technical Description).  
In usual, we make a ARF file using calibration file of contaminant, which have 
information of thickness of contaminant at the center region in chronological order. 
Furthermore, they reproduce the space distribution of contaminant with a simple function of position. 
It is very difficult, however, to completely predict the distribution of contaminant material 
because of time-varying composition, thickness, and configuration.
Especially, in recent observations, outer region of detector, and XIS0/3, 
some discrepancies between data and model under $\sim$0.6 keV 
are known even if using calibration files.

As mentioned in Section \ref{sec:data}, we have fit the spectra of all sample galaxy with ARF file 
including CALDB file contaminant and figured out there are still discrepancies for some galaxies. 
We attributed these discrepancies to difficulty to reproduce the contaminant profile 
about for the galaxies observed since the beginning of 2008. 
On the other hands, about for NGC~4472 and NGC~4649 observed in 2006, 
systematic error of contaminant estimation becomes pronounced 
because of very bright galaxies. 
We have performed spectral fit of data observed 
since the beginning of 2008 with ARF files without including the effect of the contaminant. 
Table \ref{tbl:contami} shows how we set calibration file or not, as making ARF files.
When we fit the spectra with ``varabs'' model, the parameter of C in ``varabs'' were set to free. 
Meanwhile, we set O, corresponding to the ratios of C/O is values from Table \ref{tbl:contami_ratio} 
in units of solar ratio.
The values of Table \ref{tbl:contami_ratio} are the ratios of C/O at the center of each detector 
from calibration files, ae\_xi0 (or 1 or 2 or 3)\_contami\_20091201.fits. 
In the Table \ref{tbl:contami_C}, we summarized the derived C vales of varabs model with 1T or 2T fit. 
The derived values of C is about 2/3 and 1/2 times of those values of calibration files for 
4r$_{\rm e}$ and background region, respectively.

\begin{figure*}[tbp]
 \begin{center}
\includegraphics[width=0.5\textwidth,angle=0,clip]{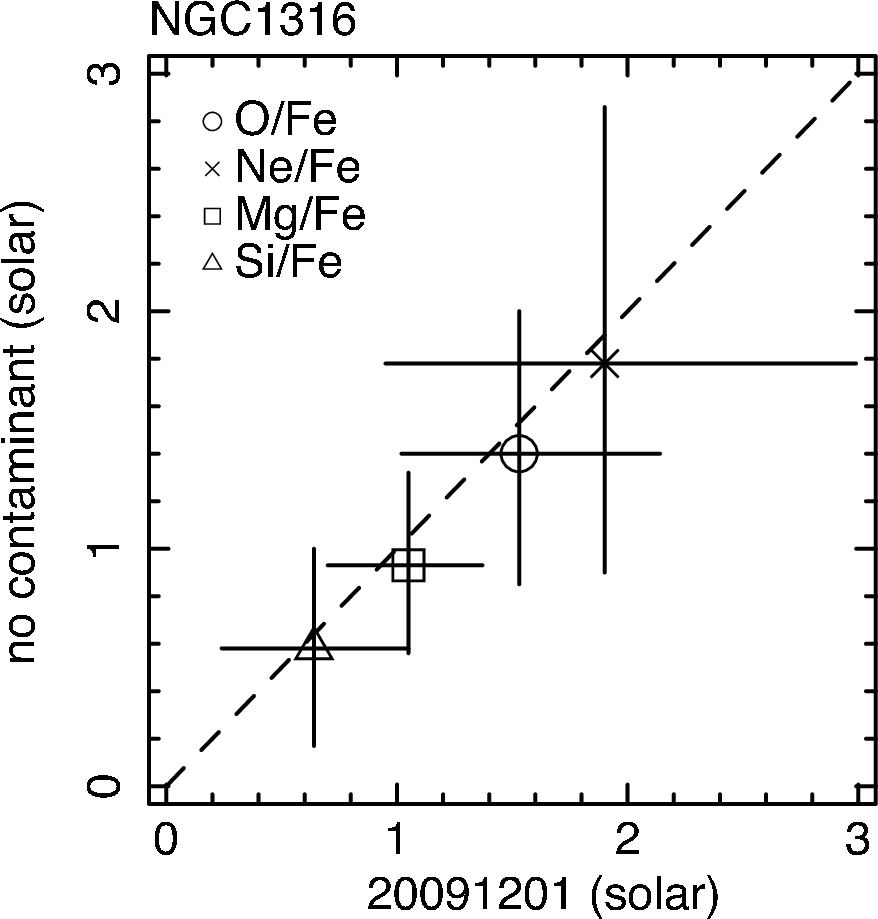}
 \end{center}
 \caption{
\rm Comparison of abundance ratios (O/Fe, Ne/Fe, Mg/Fe, and Si/Fe) with varabs model and ARF file using calibration file ae\_xi0 (or 1 or 2 or 3)\_contami\_20091201.fits. 
The dashed line indicate the equal value between two models.
}
\label{img:ngc1316_varabs}
\end{figure*}

We checked whether the results of fitting with ARF files without including the effect of contaminant 
are consistent to 
those with calibration files of contaminant, whose name are ae\_xi0 (or 1 or 2 or 3)\_contami\_20091201.fits. 
Then, NGC~1316 which observed in 2006 have been reanalyzed, using ARFs without the effect of contaminant. 
We fitted the spectra of the 4r$_{\rm e}$ and background regions simultaneously 
with same model in Section \ref{ssec:1t2t}. 
All derived parameters are consistent between the two fits. 
We also summarized the ratios of metals in Figure \ref{img:ngc1316_varabs}, these are also in good agreement.

\begin{table*}
\caption{\rm Applied Calibration file of contaminant \label{tbl:contami}}
\begin{center}
\begin{tabular}{ccccccccccc}
\tableline\tableline\noalign{\smallskip}
Galaxy   & BGD &            &          & 4r${\rm e}$  &      &                 \\
               & XIS0 & XIS1 & XIS3  & XIS0             & XIS1  & XIS3    \\ 
\noalign{\smallskip}\tableline\noalign{\smallskip}
NGC~1332  &  no & 20091201 & no & 20091201 & 20091201  & 20091201 \\
NGC~2300  &   no & 20091201 & no & 20091201 &  20091201  & 20091201 \\
NGC~4125  &  no & 20091201 & no & no & 20091201  & no\\
NGC~4382  &   no & 20091201 & 20091201 & 20091201 &  20091201  & 20091201\\
NGC~4406  &   no & no & no & no &  no  & no \\
NGC~4472  &   no & no & no & no &  no  & no \\
NGC~4649  &   no & no & no & no &  no  & no \\
NGC~4697  &  no & 20091201 & no & no & 20091201  & 20091201 \\ 
NGC~5846  &   no & 20091201 & 20091201 & 20091201 &  20091201  & 20091201\\ 
\noalign{\smallskip}\tableline\noalign{\smallskip}
\end{tabular}
\end{center}
\end{table*}

\begin{table*}
\caption{\rm The set ratios of C to O in varabs model.\label{tbl:contami_ratio}}
\begin{center}
\begin{tabular}{ccccccccccc}
\tableline\tableline\noalign{\smallskip}
Galaxy   & BGD &            &          & 4r${\rm e}$  &      &                 \\
               & XIS0 & XIS1 & XIS3  & XIS0             & XIS1  & XIS3    \\ 
\noalign{\smallskip}\tableline\noalign{\smallskip}
NGC~1332  &  7.27 & \nodata & 7.27  & \nodata & \nodata  & \nodata \\
NGC~2300  &  7.31 & \nodata & 7.31 & \nodata & \nodata  & \nodata \\
NGC~4125  & 7.33 & \nodata & 7.34  & 7.33  & \nodata  & 7.34 \\
NGC~4382  & 7.61  & \nodata & \nodata & \nodata &  \nodata  & \nodata\\
NGC~4406  & 7.37 & 7.08  & 7.10  & 7.37 & 7.08  & 7.10  \\
NGC~4472  &  9.63  &  9.64 &  9.64&  9.63  &  9.64 &  9.64 \\
NGC~4649  & 9.45  & 9.35 &  9.36  & 9.45  & 9.35 &  9.36 \\
NGC~4697  & 7.27 & \nodata & 7.27  & 7.27  & \nodata  & \nodata \\ 
NGC~5846  &  7.57 & \nodata & \nodata & \nodata &  \nodata  & \nodata\\ 
\noalign{\smallskip}\tableline\noalign{\smallskip}
\end{tabular}
\end{center}
\end{table*}

\begin{table*}
\caption{\rm The C column densities (10$^{22} \rm cm^{-2}$) in varabs model derived from the 1T or 2T model fits using APEC plasma code v2.0.1.\label{tbl:contami_C}}
\begin{center}
\begin{tabular}{ccccccccccc}
\tableline\tableline\noalign{\smallskip}
Galaxy   & BGD &            &          & 4r${\rm e}$  &      &                 \\
               & XIS0 & XIS1 & XIS3  & XIS0             & XIS1  & XIS3    \\ 
\noalign{\smallskip}\tableline\noalign{\smallskip}
NGC~1332  &  1.05$^{+0.24}_{-0.23}$ & \nodata & 1.18$^{+0.24}_{-0.23}$  & \nodata & \nodata  & \nodata \\
NGC~2300  &  1.58$^{+0.19}_{-0.18}$ & \nodata & 1.53$^{+0.18}_{-0.17}$ & \nodata & \nodata  & \nodata \\
NGC~4125  & 1.13$^{+0.22}_{-0.21}$ & \nodata & 1.42$^{+0.23}_{-0.22}$  & 1.69$^{+0.18}_{-0.18}$  & \nodata  & 1.60$^{+0.18}_{-0.18}$ \\
NGC~4382  & 0.86$^{+0.17}_{-0.17}$  & \nodata & \nodata & \nodata &  \nodata  & \nodata\\
NGC~4406  & 1.78$^{+0.12}_{-0.12}$ & 1.72$^{+0.10}_{-0.10}$  & 2.24$^{+0.12}_{-0.12}$  & 2.18$^{+0.09}_{-0.08}$ & 1.23$^{+0.21}_{-0.18}$  & 2.26$^{+0.08}_{-0.08}$  \\
NGC~4472  &  0.48$^{+0.13}_{-0.13}$  & 0.86$^{+0.11}_{-0.10}$ & 1.60$^{+0.08}_{-0.08}$ &  1.38$^{+0.08}_{-0.08}$   & 1.29$^{+0.26}_{-0.22}$  &  2.39$^{+0.08}_{-0.08}$  \\
NGC~4649  & 1.15$^{+0.18}_{-0.14}$  & 1.52$^{+0.17}_{-0.13}$ & 2.14$^{+0.19}_{-0.15}$  & 1.77$^{+0.09}_{-0.07}$  & 2.17$^{+0.07}_{-0.05}$ & 2.53$^{+0.09}_{-0.07}$ \\
NGC~4697  & 1.21$^{+0.15}_{-0.15}$ & \nodata & 1.45$^{+0.14}_{-0.13}$  & 1.84$^{+0.16}_{-0.16}$  & \nodata  & \nodata \\ 
NGC~5846  &  1.17$^{+0.07}_{-0.07}$ & \nodata & \nodata & \nodata &  \nodata  & \nodata\\ 
\noalign{\smallskip}\tableline\noalign{\smallskip}
\end{tabular}
\end{center}
\end{table*}

\section{FIT SPECTRA IN SECTION 4.1}
\label{a:analysis}

In this section, we summarized the background parameters and spectra in Section \ref{ssec:1t2t}. 
The caption of figures are same as Figure \ref{img:spec}.

\begin{table*}
\caption{\rm Best fit parameters from background region fit in the Section \ref{sssec:bgd}. Metal abundances of ``vapec'' model of NGC~1399 and NGC~4636 are summarized in Table \ref{tbl:bgdonly2}.\label{tbl:bgdonly1}}
\begin{center}
\begin{tabular}{ccccccccccc}
\tableline\tableline\noalign{\smallskip}
Galaxy   &   kT$_{\rm Galactic}$  & kT$_{\rm Galactic}$ & kT$_{\rm ETE}$ & Abundance  & ${\chi^2}$/d.o.f. \\
         &   (keV)  & (keV)  &   (keV)  & (solar) &   \\ 
\noalign{\smallskip}\tableline\noalign{\smallskip}
NGC~720  &  0.09$^{+0.03}_{-0.06}$  & 0.20$^{+0.06}_{-0.02}$ & 0.64$^{+0.09}_{-0.03}$ & 0.19$^{+0.11}_{-0.06}$ &  284/228\\
NGC~1316 &  0.09 (fix)  & 0.22$^{+0.05}_{-0.02}$ & 0.91$^{+0.04}_{-0.04}$  & 0.3 (fix) & 169/174 \\
NGC~1332 & 0.07$^{+0.02}_{-0.01}$ & 0.30$^{+0.04}_{-0.04}$  & \nodata  & \nodata  & 193/173 \\
NGC~1399 & \nodata & \nodata & 1.67$^{+0.16}_{-0.04}$  & \nodata & 698/460\\
         &  \nodata       &    \nodata         &   1.12$^{+0.12}_{-0.04}$   &  \nodata  &  \nodata    \\
NGC~1553 &  0.07 (fix) & 0.19 (fix)  & 0.75$^{+0.05}_{-0.05}$ & 0.22$^{+0.30}_{-0.09}$  & 203/175 \\
NGC~2300 & 0.13$^{+0.03}_{-0.03}$  & \nodata  &  0.93$^{+0.02}_{-0.02}$ & 0.15$^{+0.18}_{-0.03}$   &  263/202 \\
NGC~3923 & 0.11 (fix) & 0.19 (fix)  & 0.70$^{+0.03}_{-0.03}$ & 0.55$^{+2.92}_{-0.27}$  & 307/230 \\ 
NGC~4125 & 0.07$^{+0.02}_{-0.01}$  & 0.30$^{+0.02}_{-0.02}$  & \nodata  & \nodata & 211/173 \\
NGC~4382 & 0.10$^{+0.02}_{-0.01}$  & 0.28$^{+0.02}_{-0.02}$  & \nodata  & \nodata  & 197/174 \\
NGC~4406 & 0.10 (fix)  & 0.23 (fix)  & 0.97$^{+0.02}_{-0.03}$ & 0.83$^{+0.52}_{-0.23}$   &  385/349 \\
         &   \nodata         &     \nodata        & 1.52$^{+0.15}_{-0.13}$  &  0.3 (fix)             &  \nodata      \\
NGC~4472 & 0.11$^{+0.02}_{-0.02}$ & \nodata  & 1.08$^{+0.17}_{-0.03}$ & 0.73$^{+0.13}_{-0.09}$ & 380/349 \\
         &  \nodata          &   \nodata          & 1.74$^{+0.49}_{-0.12}$ &  linked                  &  \nodata \\
NGC~4552 & 0.13$^{+0.02}_{-0.04}$  & \nodata   & 2.00$^{+0.49}_{-0.32}$ & 0.3 (fix) & 271/246 \\
NGC~4636 & 0.13 (fix)  & 0.21 (fix)  & 0.89$^{+0.01}_{-0.01}$ & \nodata & 647/461 \\
NGC~4649 & 0.07 (fix) & 0.17 (fix)  & 0.98$^{+0.02}_{-0.02}$ & 0.12$^{+0.02}_{-0.01}$  & 490/351 \\
NGC~4697 & 0.11$^{+0.01}_{-0.01}$  & 0.27$^{+0.01}_{-0.02}$ & \nodata  & \nodata  &  202/173 \\
NGC~5846 & 0.12 (fix)  & 0.22$^{+0.02}_{-0.01}$  & 0.67$^{+0.09}_{-0.02}$ & 0.46$^{+0.06}_{-0.05}$  & 267/201 \\
         &   \nodata          &      \nodata                   & 1.17$^{+0.04}_{-0.03}$ &  linked              & \nodata  \\
\noalign{\smallskip}\tableline\noalign{\smallskip}
\end{tabular}
\end{center}
\end{table*}

\begin{table*}
\caption{\rm The O, Ne, Mg, Si, S, and Fe abundances in the ISM of NGC~1399 and NGC~4636 derived from background region fit in the Section \ref{sssec:bgd}.\label{tbl:bgdonly2}}
\begin{center}
\begin{tabular}{ccccccccccc}
\tableline\tableline\noalign{\smallskip}
Galaxy   &  O & Ne & Mg  & Si & S & Fe  \\
         & (solar) & (solar)  & (solar)  & (solar)  & (solar)  & (solar)    \\
\noalign{\smallskip}\tableline\noalign{\smallskip}
NGC~1399 & 0.44$\pm{0.10}$  & 2.10$^{+0.31}_{-0.30}$  & 0.78$\pm{0.08}$ & 0.82$\pm{0.05}$ & 0.82$\pm{0.07}$ & 0.77$\pm{0.04}$   \\
NGC~4636 & 0.60$^{+0.21}_{-0.20}$ & 1.22$^{+0.19}_{-0.18}$ & 0.51$^{+0.08}_{-0.07}$ & 0.38$^{+0.06}_{-0.05}$ & =Si & 0.36$^{+0.03}_{-0.03}$   \\
\noalign{\smallskip}\tableline\noalign{\smallskip}
\end{tabular}
\end{center}
\end{table*}

\begin{figure*}
\centering
\includegraphics[width=0.8\textwidth,angle=0,clip]{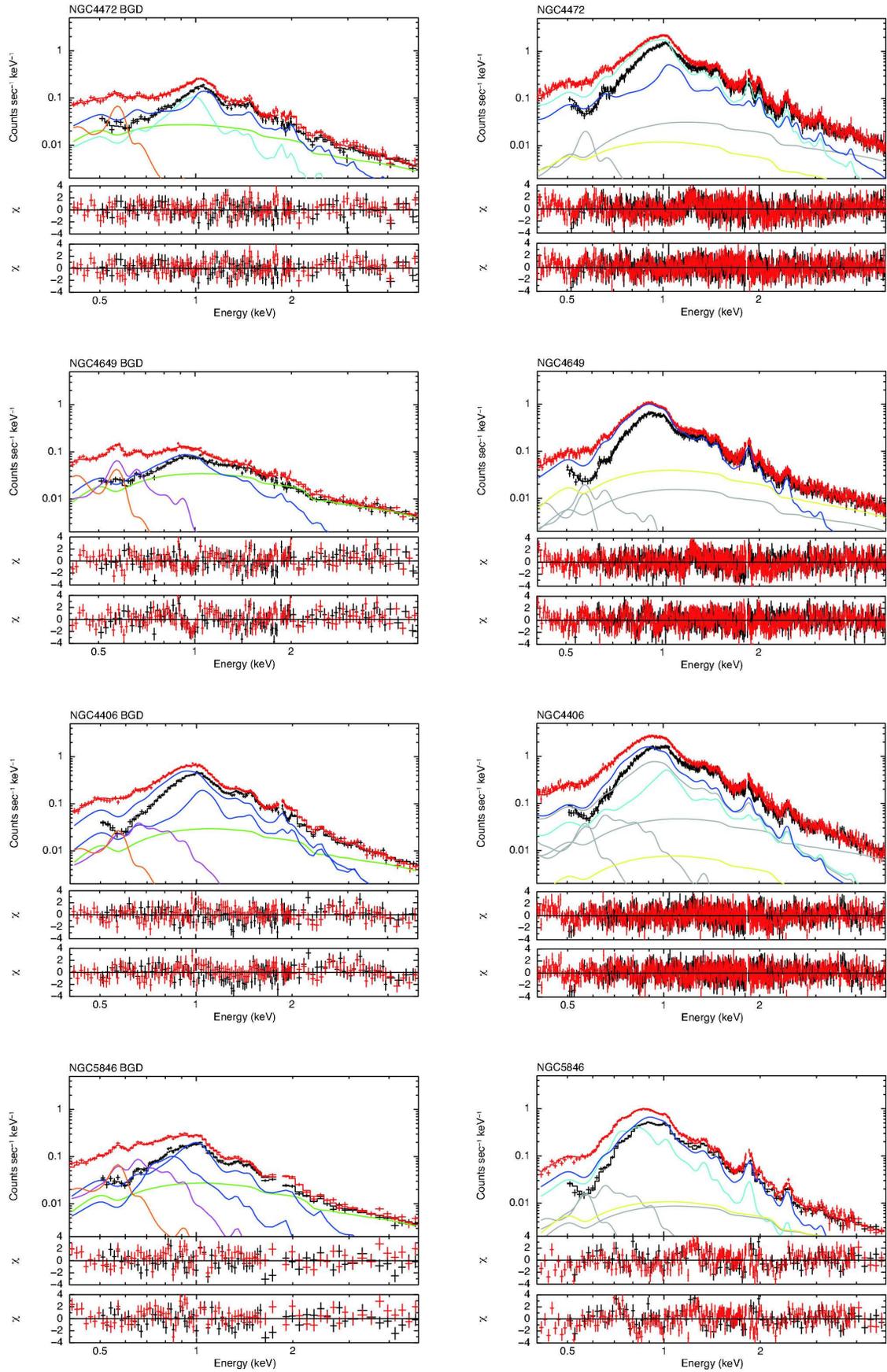}
\caption{The same as Figure \ref{img:spec} for the galaxies
are not shown in the Figure \ref{img:spec}.
}
\label{aimg:spec1}
\end{figure*}

\addtocounter{figure}{-1}
\begin{figure*}
\centering
\includegraphics[width=0.8\textwidth,angle=0,clip]{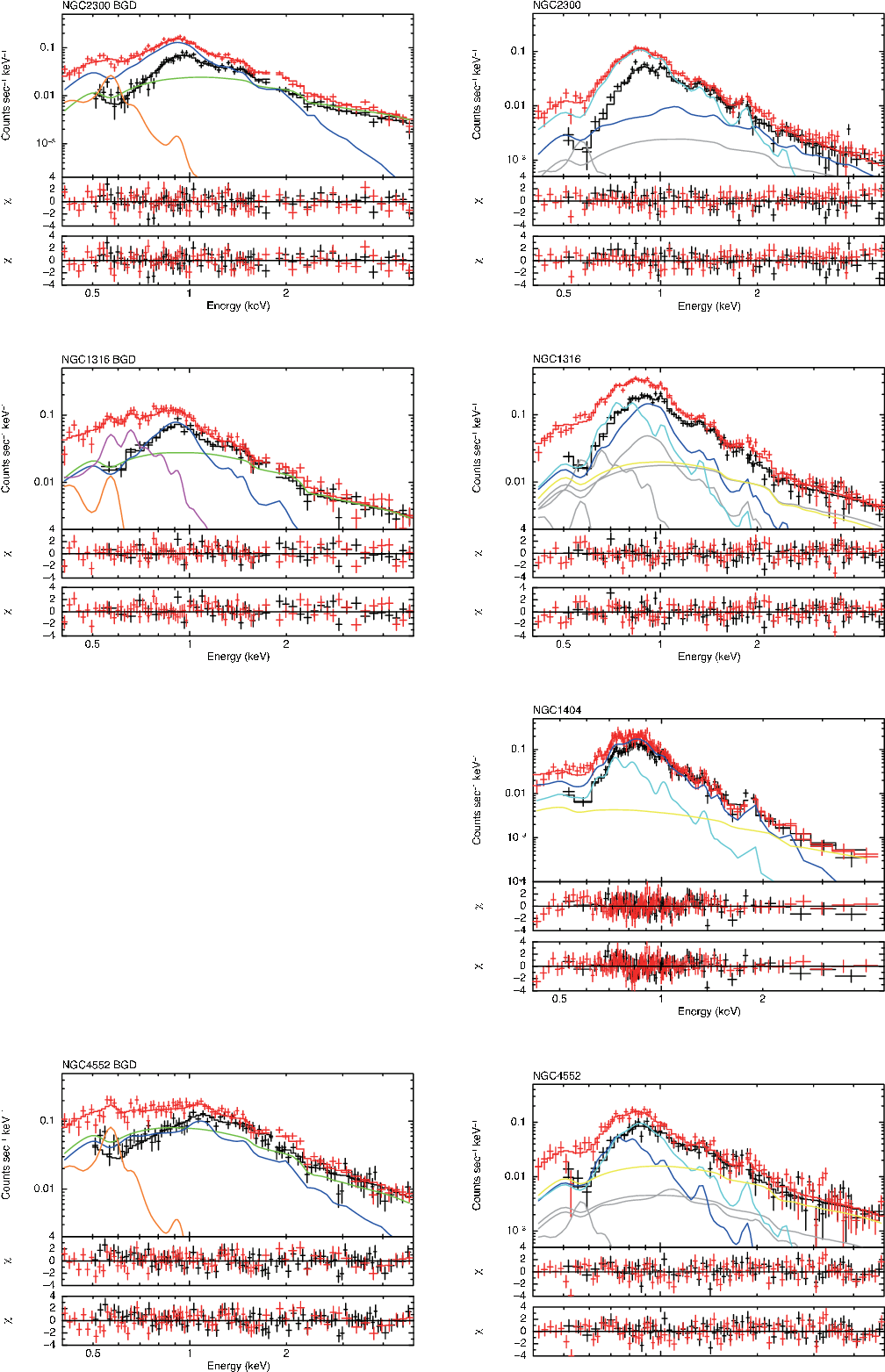}
\caption{
        {\it Continued.}
}
\label{figure:pc_result:b}
\end{figure*}

\addtocounter{figure}{-1}
\begin{figure*}
\centering
\includegraphics[width=0.8\textwidth,angle=0,clip]{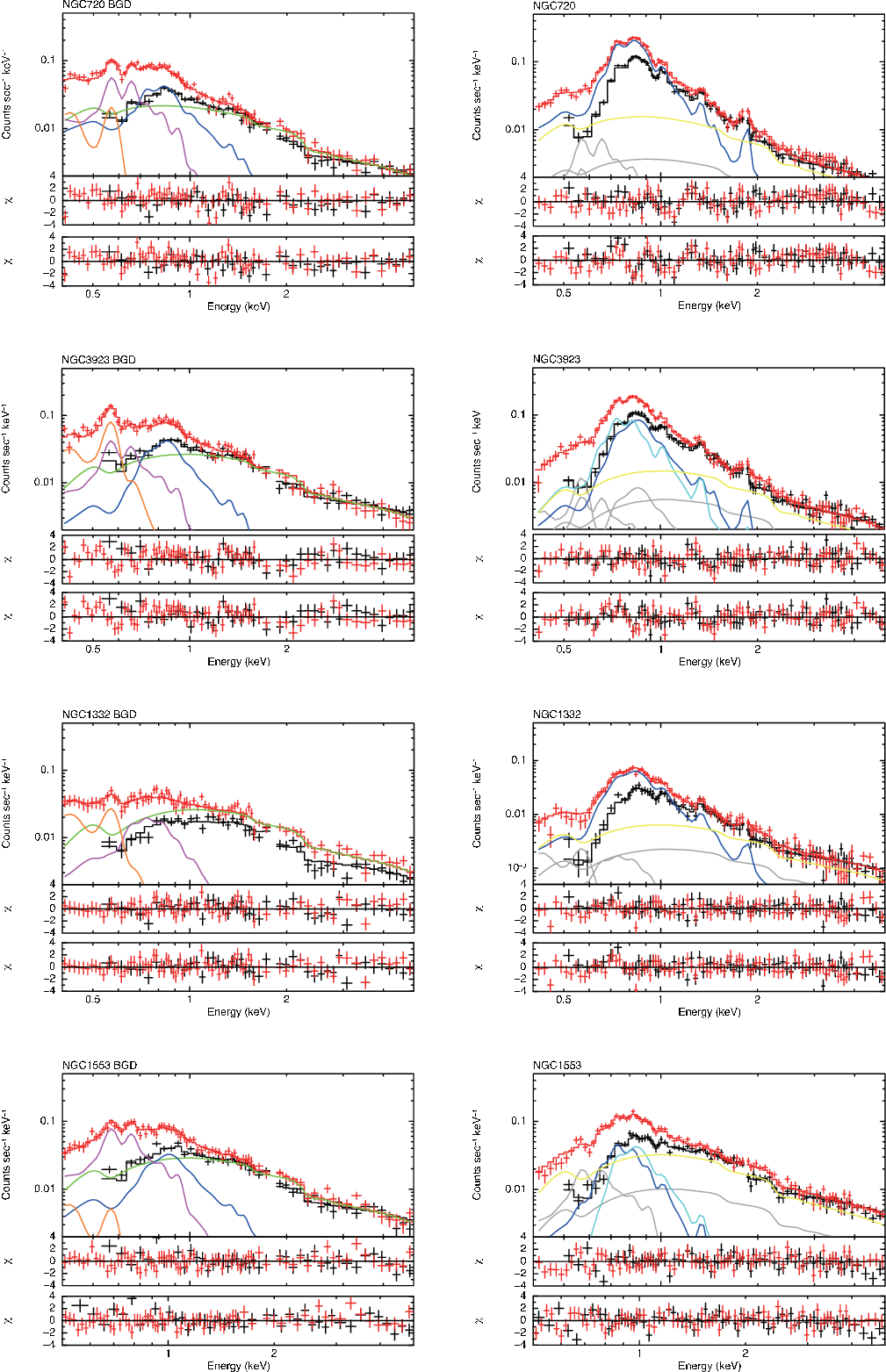}
\caption{
        {\it Continued.}
}
\label{figure:pc_result:b}
\end{figure*}

\addtocounter{figure}{-1}
\begin{figure*}
\centering
\includegraphics[width=0.8\textwidth,angle=0,clip]{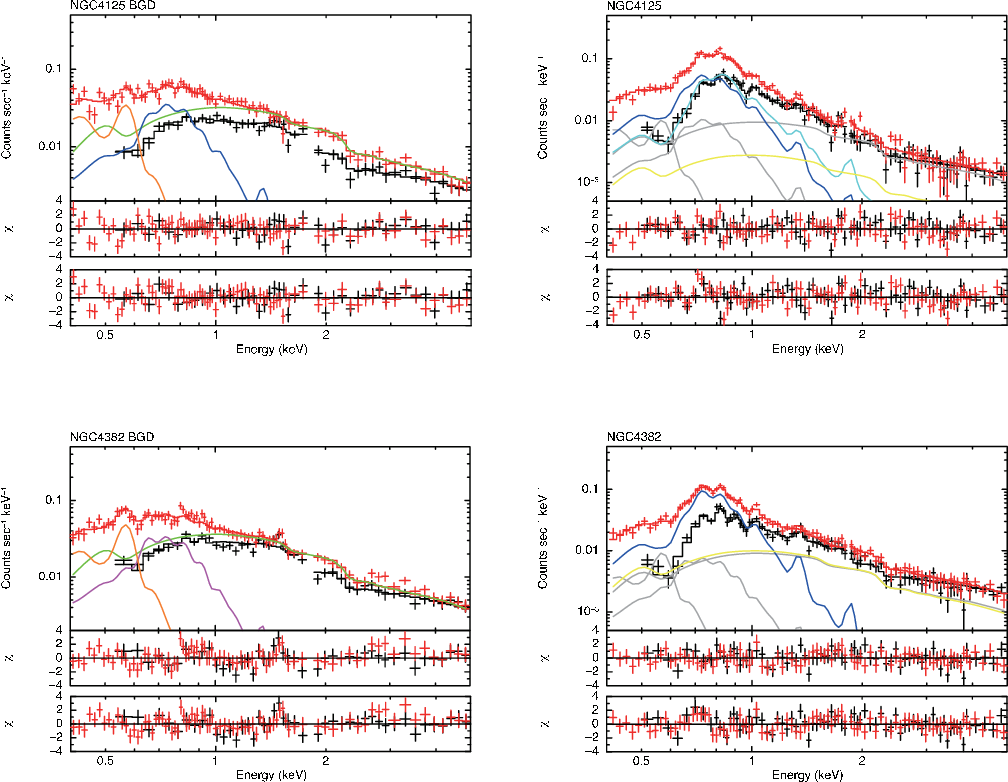}
\caption{
        {\it Continued.}
}
\label{figure:pc_result:b}
\end{figure*}


\end{document}